\documentclass{nature}

\usepackage[pdftex]{graphicx}
\usepackage[english]{babel}  %<--hyphenation
\usepackage{amssymb}         %<--symbols
\usepackage[fleqn]{amsmath}         %<--fancy equations
\usepackage{caption}
\usepackage{enumerate}
\usepackage{subfigure}
\usepackage{authblk}
\usepackage[T1]{fontenc}
\usepackage[table]{xcolor}
\usepackage{makeidx}
\usepackage[normalem]{ulem}
\usepackage{color}
\usepackage{bm}% bold math
\usepackage{hyperref}   % use for hypertext links, including those to external documents and URLs

\usepackage{booktabs} % for fancy tables
\usepackage{tabularx} % stretched tables
\usepackage{wrapfig}

\title{The relationship between human mobility and viral transmissibility during the COVID-19 epidemics in Italy}

\author[1]{Paolo Cintia}
\author[2, $\dagger$]{Luca Pappalardo}
\author[2]{Salvatore Rinzivillo}
\author[2]{Daniele Fadda}
\author[3]{Tobia Boschi}
\author[2]{Fosca Giannotti}
\author[3, 4]{Francesca Chiaromonte}
\author[6]{Pietro Bonato}
\author[6]{Francesco Fabbri}
\author[6]{Francesco Penone}
\author[6]{Marcello Savarese}
\author[5]{Francesco Calabrese}
\author[7]{Flavia Riccardo}
\author[7]{Antonino Bella}
\author[7]{Xanthi Andrianou}
\author[7]{Martina Del Manso}
\author[7]{Massimo Fabiani}
\author[7]{Stefania Bellino}
\author[7]{Stefano Boros}
\author[7]{Alberto Mateo Urdiales}
\author[7]{Maria Fenicia Vescio}
\author[7]{Silvio Brusaferro}
\author[7]{Giovanni Rezza}
\author[7]{Patrizio Pezzotti}
\author[8]{Marco Ajelli}
\author[8]{Giorgio Guzzetta}
\author[8]{Valentina Marziano}
\author[8]{Stefano Merler}
\author[8]{Piero Poletti}
\author[8]{Filippo Trentini}
\author[9]{Paolo Vineis}
\author[1, $\dagger$]{Dino Pedreschi}

\affil[1]{Department of Computer Science, University of Pisa, Italy}
\affil[2]{Institute of Information Science and Technologies (ISTI), National Research Council (CNR), Italy}
\affil[3]{Department of Statistics and Huck Institutes of the Life Sciences, Penn State University, USA}
\affil[4]{Institute of Economics and EMbeDS, Sant’Anna School of Advanced Studies, Italy}
\affil[5]{Vodafone Analytics}
\affil[6]{WindTre}
\affil[7]{Istituto Superiore di Sanità (ISS)}
\affil[8]{Fondazione Bruno Kessler (FBK)}
\affil[9]{Imperial College London, School of Public Health}
\affil[$\dagger$]{To whom correspondence should be addressed: luca.pappalardo@isti.cnr.it, dino.pedreschi@unipi.it}

\begin{document}

\maketitle

\vspace{1cm}

\begin{abstract}
In 2020, countries affected by the COVID-19 pandemic implemented various non-pharmaceutical interventions to contrast the spread of the virus and its impact on their healthcare systems and economies.
Using Italian data at different geographic scales, we investigate the relationship between human mobility, which subsumes many facets of the population’s response to the changing situation, and the spread of COVID-19. 
Leveraging mobile phone data from February through September 2020, we find a striking relationship between the decrease in mobility flows and the net reproduction number.
We find that the time needed to switch off mobility and bring the net reproduction number below the critical threshold of $1$ is about one week.
Moreover, we observe a strong relationship between the number of days spent above such threshold before the lockdown-induced drop in mobility flows and the total number of infections per 100k inhabitants. 
Estimating the statistical effect of mobility flows on the net reproduction number over time, we document a 2-week lag positive association, strong in March and April, and weaker but still significant in June.

Our study demonstrates the value of big mobility data to monitor the epidemic and inform control interventions during its unfolding.

\end{abstract}

%% Here is the endmatter stuff: Supplementary Info, etc.
%% Use \item's to separate, default label is "Acknowledgements"

\section{Introduction}

During the COVID-19 pandemic, countries put in place a wide variety of non-pharmaceutical interventions \cite{perra2020nonpharmaceutical, bo2020effectiveness,  haug2020ranking}, such as closures of activities and travel restrictions,
%to the aim of 
aimed at containing the spread of the virus, mitigating the impact on the healthcare system, and saving lives. 
Accordingly, citizens adjusted their behavior to the restrictions and the recommendations for personal protection, such as wearing face masks and maintaining social distance. 
Understanding and monitoring the pandemic's evolution in 
%such 
a continuously changing 
%situation 
context of policy interventions and behavioral shifts 
%are very 
is challenging
%, albeit 
and yet crucial for policymakers -- who need to evaluate the effects of the implemented measures and plan future restrictions or relaxations as the situation unfolds. 
While different social, economic, demographic, cultural, and psychological variables are at play in the complex and quickly evolving
%pandemic 
scenario of a pandemic, one physical quantity subsumes many facets of the population's response: 
%to the changing situation: 
{\em mobility}, i.e.~the number of trips, at any scale, that people take to perform their daily activities. 
Human mobility is a powerful, objective proxy 
%of 
for people's behavior and its adaptation 
%and 
in response to 
%the 
different stimuli during 
%the 
a pandemic, and we have a reliable tool for observing and quantifying human mobility with precision at country scale: mobile phone data \cite{blondel2015survey, barbosa2018human, luca2020deep, giannotti2008mobility, wang2018applying}. 

More than a decade of experience in acquisition and analysis of geolocated, spatio-temporal traces left behind by smartphone users has already shown how models based on such data can account for different aspects of our society, also beyond mere mobility \cite{giannotti2008mobility, gonzalez2008understanding, pappalardo2015returners, alessandretti2018evidence, simini2012universal, pappalardo2016analytical, eagle2009eigenbehaviors, song2010limits, barbosa2020uncovering}. 
Mobile phone records provide an unprecedented opportunity to track human displacements \cite{blondel2015survey, barbosa2018human, giannotti2008mobility, luca2020deep}, allowing one to estimate crucial societal aspects like population density \cite{gabrielli2015city, deville2014dynamic,  douglass2015high}, mobility patterns and flows \cite{gonzalez2008understanding, pappalardo2015returners, hankaew2019inferring, balzotti2018understanding, bonnel2018origin}, well-being \cite{pappalardo2016analytical, liang2020using, voukelatou2020measuring, eagle2010network, frias2012relationship, blumenstock2015predicting}, and migrations \cite{lai2019exploring, chi2020general, hankaew2019inferring, blumenstock2012inferring, hughes2016inferring}. 
Mobile phone data analytics can be properly designed to preserve privacy and anonymity \cite{demontoye2018privacy, pellungrini2017data, fiore2019privacy, demontjoye2013unique, giannotti2008mobility}, and is advocated also as a resource for supporting public health actions across the phases of the COVID-19 pandemic \cite{oliver2020mobile, buckee2020aggregated}.
Motivated by the potential of mobile phone data in capturing the geographical spread of epidemics \cite{finger2016mobile, tizzoni2014use, wesolowski2012quantifying, bengtsson2015using}, researchers and governments have started to collaborate with mobile network operators to estimate the effectiveness of control measures in several countries \cite{kraemer2020effect, bonato2020mobile, pullano2020population, lai2020effect, liautaud2020fever, badr2020association, coven2020disparities, gozzi2020estimating, bakker2020effect,  jia2020population, kissler2020reductions, gibbs2020changing, kang2020multiscale, pepe2020covid}.

In this paper, we investigate in depth the relationship between human mobility and the spread of COVID-19 at different geographic scales, leveraging nation-wide Italian mobile phone data provided by two major telecom operators, Vodafone and WindTre, spanning 9 months since the start the outbreak of the virus.\footnote{From February till September 2020. We are currently extending the analysis for the remaining months of 2020.} 
We compare the evolution of daily mobility flows and the evolution of viral transmissibility, measured by the net reproduction number (mean number of secondary infections generated by one primary infector) from January through September 2020 -- a period in which several control interventions and human behavioral adaptations took place across the country. 
As a key result, we find a striking relationship between the decrease in mobility flows and the decrease in net reproduction number in all Italian regions between March 11th and March 16th, when the country entered the national lockdown. 
We can quantify the time needed to switch off mobility and the time required to bring the net reproduction number below the critical threshold of 1 (one week). 

We also find a strong relationship between the number of days spent above the epidemic threshold before the lockdown-induced drop in mobility flows and the total number of confirmed SARS-CoV-2 infections per 100k inhabitants. 
While many other factors may have played a role in containing the contagion, including other non-pharmaceutical interventions, this observation provides strong evidence supporting the effectiveness of mobility restrictions.
Finally, we use Functional Data Analysis tools 
to estimate the statistical effect of mobility flows on the net reproduction number over time. 
We document a 2-week lag positive association that was very strong in March and April and weaker but still significant in June.

Understanding the relationship between human mobility patterns and the spread of COVID-19 is crucial, in particular, for restarting social and economic activities, which were limited or put in stand-by during national and regional lockdowns, as well as to monitor the risk of a resurgence during lockdown exits. 
Our study, developed within the context of the Italian data-driven task force to support policy making during the COVID-19 epidemics, demonstrates the value of "big" mobility data for monitoring the epidemic and inform control interventions during its unfolding.

\section{Data sets}
\subsection{Mobility data.}

%Mobile phone data have proven to be a useful data source to track human mobility \cite{blondel2015survey,  barbosa2018human, gonzalez2008understanding, pappalardo2015returners}, and thus a tool for monitoring the effectiveness of control measures such as movement restrictions and physical distancing \cite{oliver2020mobile, buckee2020aggregated, lai2019measuring, kraemer2020effect, lai2020effect, pullano2020population, bonato2020mobile, badr2020association} and their relationships with socio-demographic characteristics \cite{coven2020disparities, bonaccorsi2020economic, weill2020social, gozzi2020estimating, kissler2020reductions}.

%The 
We use data derived from the records the two largest mobile phone operators in Italy: WindTre and Vodafone, which cover together approximately two thirds of the Italian mobile phone users.%\footnote{Since we do not find any significant difference in the results, in this paper we present the results for Vodafone only.}
The basic geographical unit is the phone cell, defined as the area covered by a single antenna, i.e., the device that captures mobile radio signals and keeps the user connected with the phone network.
Multiple antennas are usually mounted on the same tower, each covering a different direction. 
The tower's geographic position and the antenna's direction allow one to infer the extension of the corresponding phone cell. 
A user's position is approximated by the antenna serving the connection, whose extension is relatively small in urban contexts (in the order of 100m $\times$ 100m) and larger in rural areas (in the order of 1km $\times$ 1km or more). 
A user's mobility is reconstructed as the sequence of towers their phone is connected to in time. A connection (event) is created every time the user calls or makes an upload/download operation from the Internet.
%For a 4G connection, users have a few hundred events per day, on average.
%Each event is described by the antenna identifier, the phone identifiers, and the amount of kilobytes downloaded during the process.
%The identity of the users is replaced by artificial identifiers, a pseudonymization procedure that is a first important step 
%(mentioned in Article 6(4) and Article 25(1) of the GDPR, the EU General Data Protection Regulation) to provide anonymity \cite{demontjoye2013unique, demontoye2018privacy, pellungrini2017data, fiore2019privacy, pellungrini2020modeling}. 
%In this work, we consider 
We gather such data for the period from February 01st, 2020, to September 30th, 2020.

\begin{figure}
\centering
\includegraphics[width=0.8\linewidth]{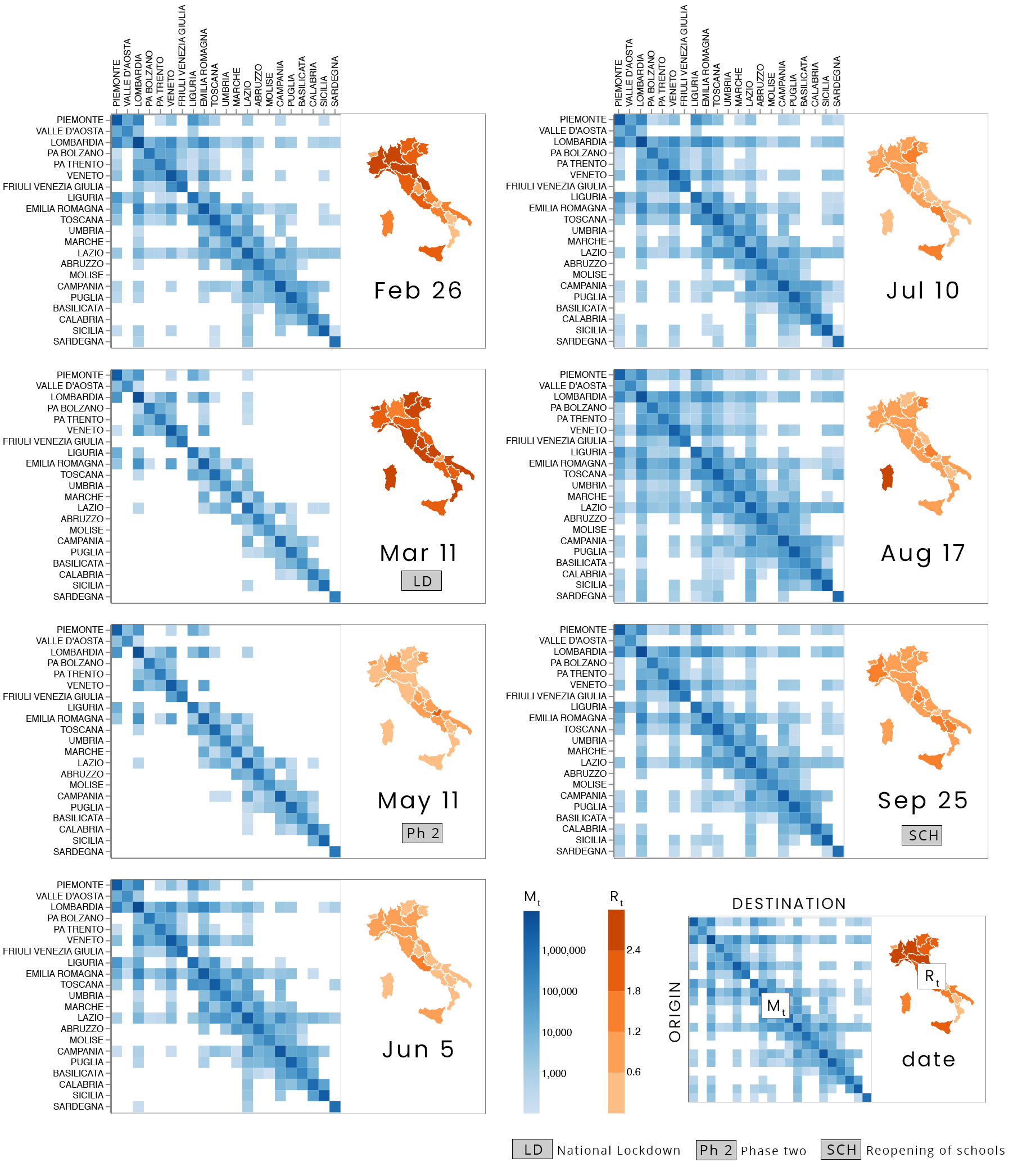}
    \caption{\footnotesize Evolution of mobility flows between origin and destination regions, and regional $R_t$ levels, for a selection of seven days in the period February 2020 - September 2020. 
    The adjacency matrix on the left of each panel depicts mobility flows, with a color intensity proportional to the number of trips between origin and destination. 
    Both flows and heterogeneity in destinations decreased during lockdown or phase 2 (e.g., March 11, May 11), with a tendency for flows to persist within neighboring geographical clusters. 
    The choropleth maps of Italy on the right  of each panel show the intensity of $R_t$ for each region.
}
    \label{fig:OD_mob}
\end{figure}

%Calculation of stay locations is the first step of the analytical pipeline, performed through network data aggregation at network towers level. 
%To filter out noise and catch only significant events, 
%to catch significant events, estimate how long a user stays in the union of the areas covered by all towers located in the same municipality.
%to generate daily trips and build Origin-Destination (OD) matrices used to measure mobility patterns across municipalities, provinces and regions. 
%The second step is to enrich stay locations: we enriched each stop with sex and age information of the SIM owner, so that an aggregated level it is possible to drill-down the analysis to obtain more useful insights able to provide the right answers to authorities questions, e.g. young people not respecting the restrictions.
We reconstruct daily Origin-Destination matrices (ODs) \cite{iqbal2014development, bachir2019inferring, bonnel2018origin} in which the flow between two municipalities $i$ and $j$ is the number of individuals that moved from $i$ to $j$ in a given day.
To reduce noise, we consider only movements for which the individual remains in the destination municipality for at least $30$ minutes.
For privacy-preserving purposes, we remove from the OD matrices flows with less than 15 trips between two municipalities. 
To match the available COVID-19 data, we aggregate the municipality-to-municipality ODs into province-to-province and region-to-region ODs, in which each node represents an Italian province or region. 
In particular, for each day, we compute both the in-flows, which capture the total number of people moving to a province/region from any other province/region, and the self-flows, which capture trips between municipalities of the same province/region. 
We then sum in-flows and self-flows to form  a daily time series $M_t$ for each province/region.
%, the sum of the in-flows, and the province/region's self-flows. 
%We normalize $M_t$ by multiplying its values by coefficients provided by the mobile phone operator, which indicate an estimation of market share for every municipality. 
Figure \ref{fig:OD_mob} shows the structure of the region-to-region OD matrices for different days in the period of observation.
%After this transformation, we have an estimation of the real size of the mobility flow between province/region.
As expected, we found no statistically significant differences between the OD matrices from the two mobile phone operators, to the purpose of the analyses presented in this paper. This observation brings robustness to the results, and allows use to use any of the two sources.
%{\color{red} We also used other metrics, see Supplementary X.}

\subsection{Epidemiological data.}

%We compare the evolution of $M_t$ with the evolution of the daily disease transmissibility in Italian regions in terms of the net reproduction number 
For each  Italian province/region, we consider the daily time series $R_t$  of the net reproduction number, which represents the mean number of secondary infections generated by one primary infector.
%, in the presence of control interventions and human behavioural adaptations. 
When $R_t$ decreases below the epidemic threshold of $1$, the number of new infections begins to decline. 
%The estimates of 
$R_t$ 
%were computed 
was estimated from the daily time series of new cases by date of symptom onset. 
Case-based surveillance data used for estimating $R_t$ were collected by regional health authorities and collated by the Istituto Superiore di Sanità (ISS) using a secure online platform, according to a progressively harmonized track-record. 
This data includes, among other information, the place of residence, the date of symptom onset, and the date of first hospital admission for laboratory-confirmed COVID-19 cases \cite{riccardo2020epidemiological}. 
The distribution of $R_t$ was estimated 
%by applying 
with a well-established statistical method \cite{who2014ebola, cori2013new, liu2018measurability}
%, which is 
based on the knowledge of the distribution of the generation time and on the time series of cases. 
In particular, the posterior distribution of $R_t$ for any time 
%point 
$t$ was estimated by applying 
%the 
Metropolis-Hastings MCMC sampling to 
%a 
the likelihood function 
%defined as follows:
\begin{equation}
L = \prod_{t=1}^T P \Big( C(t); R_t \sum_{s=1}^T \phi(s)C(t-s) \Big)
\end{equation}
where $P(x;\lambda)$ is the probability mass function of a Poisson distribution 
%(i.e., the probability of observing events if these events occur with rate)
(i.e., the probability of observing $x$ events if these events occur with rate $\lambda$); $C(t)$ is the daily number of new cases having 
%symptom 
symptoms onset at time $t$; $R_t$ is the net reproduction number at time $t$ 
%to be estimated; 
(the quantity to be estimated); and $\phi(s)$ is the probability distribution density of the generation time evaluated at time $s$.

As a proxy for the distribution of the generation time, we used the distribution of the serial interval, estimated from the analysis of contact tracing data in Lombardy \cite{riccardo2020epidemiological}, i.e., a gamma function with shape 1.87 and rate 0.28, having a mean of 6.6 days. This estimate is within the range of other available estimates for SARS-CoV-2 infections, i.e., between 4 and 7.5 days \cite{nishiura2020serial, wu2020estimating, li2020early}.
%Figure \ref{fig:OD_mob} shows the value of $R_t$ for Italian regions in different days of the observation period.
%
The number of reported SARS-COV-2 infections is provided by Protezione Civile, the Italian public institution in charge of monitoring the COVID-19 emergency. They collect data from every Italian administrative region and make them available on a public repository. 
For each province/region, we focus on the number of new positive cases per day. Specifically, given a day $g$, we average the values over the four days before and the four days after $g$.
Figure \ref{fig:OD_mob} shows the value of $R_t$ for Italian regions in different days of the observation period.

\section{Relationship between $M_t$ and $R_t$}

\begin{figure}
\centering
\includegraphics[width=0.65\linewidth]{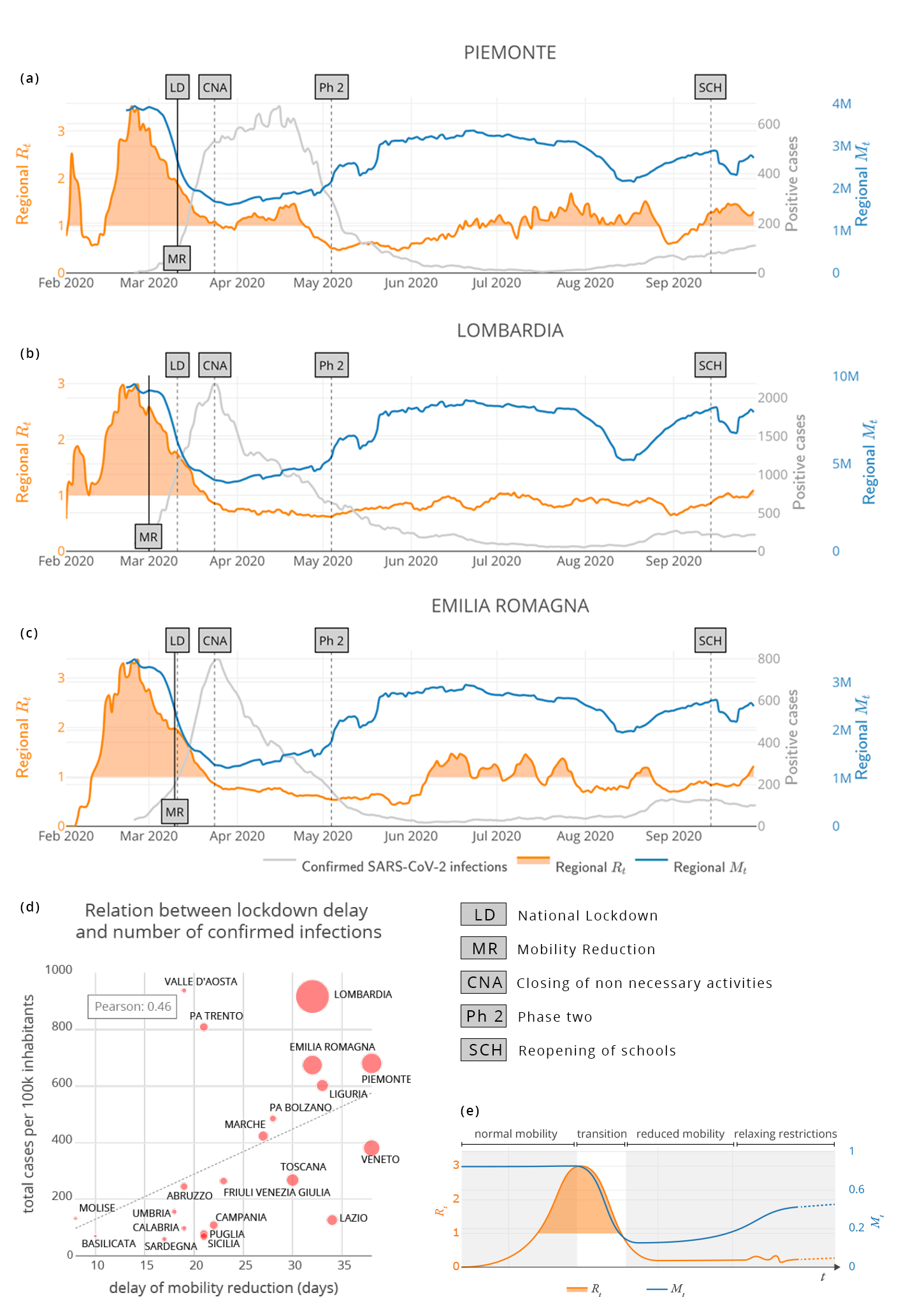}
\caption{\scriptsize Evolution of $M_t$ (blue), $R_t$ (orange, 7-day moving average) and number of confirmed infections (grey; 9-day moving average) between Feb - Sep 2020 for Piemonte (a), Lombardia (b), Emilia-Romagna (c).
Orange-shaded areas mark periods where $R_t>1$.
Vertical dashed lines indicate: beginning of national lockdown (LD, March 9th, 2020), closing of non-essential economic activities (CNA, March 23th, 2020), partial restart of economic activities and within-region mobility (“Ph 2”, May 4th, 2020), and beginning of the school year (SCH). 
%The beginning of mobility reduction (MR) does not necessarily coincide with the national lockdown (e.g., Lombardia). 
In panel (d), regions are represented in terms of the number of days between the first occurrence of $R_t > 1$ and the beginning of the national lockdown (horizontal axis), and the cumulative incidence of confirmed SARS-CoV-2 infections per 100k inhabitants as of May 15th, 2020 (vertical axis). 
The size of the circles is proportional to the total number of positive cases in the period. 
A linear least square fit (dotted line) indicates a positive relationship. 
Panel (e) schematizes the epi-mob pattern: during the first week of lockdown, the two curves describing mobility flows and net reproduction number gracefully overlap. 
At the end of this week, the country reaches a new mobility regime at $\approx$40\% of the pre-lockdown level. 
}
\label{fig:betas_mob}
\end{figure}

Figure~\ref{fig:betas_mob} shows $M_t$ (blue curves), $R_t$ (orange curves) and the number of positive cases (grey curves) for three Italian regions (similar plots for all other regions are provided in Supplementary Figure S1).
In all regions, $M_t$ decreases sharply soon after the first national lockdown (March 11th) and stabilizes on a new, reduced level after about one week. 
Subsequent restriction ordinances, such as the closing of non-essential economic activities on March 17th , have a minor impact on the reduction of $M_t$. 
In almost all regions, $M_t$ increases at the start of ``phase 2" (partial lifting of the lockdown) on May 4th. This behaviour is particularly pronounced for Emilia-Romagna (Figure \ref{fig:betas_mob}c), Toscana, Puglia and Lazio (Supplementary Figure S1 j,m,q). 
Interestingly, we find a slight increase of $M_t$ approaching May 4th. 
%the starting of ``phase 2'' during which a wider range of movement within regions has been allowed by the government. We interpret this result as 
This may be due to a progressive, although slight, relaxation of compliance with the mobility limitations imposed by the lockdown. 
The case of Molise is different and particularly compelling: it is indeed the only region for which $M_t$ decreases after May 4th (Supplementary Figure S1 o). 
This may be due to news media coverage of a funeral on April 30th, attended by a large number of people, which resulted in a large local outbreak. 
The news may have induced parts of the population of Molise to self-restrict movements during the following days. 
$M_t$ returns to the original level between May and June, around two weeks after the starting of phase 2.

Although the date when $R_t$ surpasses $1$ for the first time varies from region to region, for most of the regions $R_t$ decreases concurrently with $M_t$ due to the beginning of the national lockdown (see Figure 2, Supplementary Figure S2), highlighting the importance of governmental interventions to reduce mobility.
Note that, for most of the regions, $R_t$ 
%starts taking 
reaches values lower than $1$ 
%since 
starting around March 16th, when $M_t$ stabilizes on the new, reduced level until May (Supplementary Figure S1).
$M_t$ increases between May and June. 
Yet, in this period, 
%still 
$R_t$ continues to decrease.
This may be due to human mobility not coming back to the usual volume and connectivity, as well as to other ordinances by local and national  governments related to the wearing of masks and gloves in public areas, social distancing and ban on gatherings -- and possibly to other factors to be further investigated.
While $M_t$ remains stable until mid August, in many regions $R_t$ starts slightly increasing again from June 
%on for many regions, though 
onward, though with different patterns in different regions.
This result indicates that 
%realising 
releasing mobility restrictions for a long enough time boosts the circulation of the virus, highlighting \emph{a fortiori} the relationships between mobility and viral transmissibility.

We go one step further in our analysis of the 
%relationships 
relationship between mobility flows and contagion by computing two quantities for each region: \emph{(i)} the delay in mobility reduction, i.e., the number of days in which $R_t > 1$ before the regional $M_t$ decreases by at least 20\% w.r.t. the usual (pre-epidemics) weekly mobility, observed over the first two weeks of February; and \emph{(ii)} the total number of reported SARS-CoV-2 infections per 100k inhabitants in the region (as of May 15th, 2020, when national mobility restrictions were released). 

The date of mobility reduction below 20\% for each region is indicated as the MR black vertical line in 
%the time series of
Figures~\ref{fig:betas_mob}(a), (b) and (c). 
As shown in Figure~\ref{fig:betas_mob}(d),
%correlates 
the two quantities 
%for all regions, where the size of the circle of each region is proportional to the total number of reported infections in the observation period. 
%The positive correlation between the two quantities is robust
are positively and significantly correlated (Pearson coefficient= 0.46, $p<0.05$, $r^2 = 0.21$) 
-- suggesting that larger delays could have induced heavier spreading of the virus. This is  strong evidence that timely lockdowns are instrumental for better containment of the contagion.

The delay in mobility reduction in Lombardy was 32 days, leading to the highest number of positive cases per inhabitant in Italy. 
Similarly, for other regions severely affected by the virus, such as Liguria, Emilia-Romagna and Piemonte, the delays in mobility reduction were 32 and 38 days, respectively.
The north-central and north-western regions all lie above the dashed regression line in the top right portion of the scatter plot in Figure~\ref{fig:betas_mob}(d). 
Regions below this line in the bottom right portion of the plot, such as Veneto, Lazio and Tuscany,  were more effective in containing the contagion, despite delays in mobility reduction of $30$ days or more. 
This fact may be explained by several factors, including the effectiveness of the epidemic surveillance, the intensity of the testing and tracing strategy adopted, the capacity of outbreak containment, and also the absolute number of cases when $R_t$ jumps above $1$. 
On the other hand, southern regions all had smaller delays in mobility reduction -- in the range of $10$ (Molise, Basilicata) to $20$ (Campania, Puglia, Sicily, Calabria) days. 
%the mobility reduction started with around 10 days of delay (Molise, Basilicata) or around 20 days (Campania, Puglia, Sicily, Calabria), which presumably was effective 
This appears to have helped in containing the spread of the virus; all southern regions (with the only exception of Molise) are below the regression line in the bottom left portion of the plot, with low numbers of infections per 100K inhabitants. 
In summary, the regions that appear to have benefited the most from the lockdown are those where the reduction in mobility occurred in a more timely fashion.
%Central-southern regions are the ones who benefited the most from the lockdown, presumably because it started more timely. This brings further evidence of the effectiveness of the lockdown.

\section{Lags and Function-on-function regression }

% \begin{figure}
% % \begin{center}
% % {\small \bf unshifted curves 2nd solution}
% % \end{center}
% \centering
% \includegraphics[width=0.32\linewidth]{./img/fda_img/shifts.pdf}
% \includegraphics[width=0.32\linewidth]{./img/fda_img/rt_shifted.pdf}
% \includegraphics[width=0.32\linewidth]{./img/fda_img/trips_shifted.pdf}
% \caption{{\red write caption}
% }
% \label{fig:shifted_curves}
% \end{figure}

% \begin{figure}
% % \begin{center}
% % {\small \bf unshifted curves 2nd solution}
% % \end{center}
% \centering
% \includegraphics[width=0.35\linewidth]{./img/fda_img/beta_surf.pdf}
% \hspace{1cm}
% \includegraphics[width=0.45\linewidth]{./img/fda_img/beta_curve.pdf}
% \caption{{\red write caption}
% }
% \label{fig:betas_mob}
% \end{figure}

Another way to characterize the relationship between
$M_t$ and $R_t$ is to employ statistical tools from
%We use 
Functional Data Analysis (FDA). These allow us to efficiently exploit the longitudinal information at our disposal, evaluating lags and regressing transmissibility onto mobility.  
%to model the relationship between $R_t$ and $M_t$.
%, representing them as curves defined on a continuous time domain.
%Specifically, 
First, we smooth the discrete observations of $R_t$ and $M_t$ using B-splines basis functions of order 4 \cite{ramsay2005}.
To remove 
%the 
daily fluctuations
%' effects and the 
and weekly trends, we use 32 basis functions -- one per week -- and add a roughness penalty on the second derivative of the curves. 
%second derivatives. Since the $M_t$ data collection is unreliable from July 18th to July 30th, from August 18th to August 31th and from September 20th to September 23rd, we do not place any basis functions in these time periods while smoothing $M_t$. 
We select the smoothing parameter minimizing the average generalized cross-validation error across all the curves 
\cite{craven1978smoothing}.
Panels (a) and (b) in Supplementary Figure S2 show smoothed $R_t$ and $M_t$ curves for the twenty Italian regions, respectively. Based on the averages of these two sets of curves (black solid lines), the delay between the 
$M_t$ peak (late February) and 
the $R_t$ peak (early March) is approximately $2$ weeks -- 13 days.
%This suggests that $M_t$ at time $t$ affect $R_t$ at time $t + \red{13}$, i.e., there is a two weeks delay between $M_t$ and $R_t$.
A similar lag is supported considering
%To give further evidence in support of this claim, we compute the curves'
the projections of the curves on their first functional covariance component (FCC) \cite{boschi2018covariance}. 
The first FCC identifies a reduced functional space which captures the most important mode of covariation between $R_t$ and $M_t$, and allows us to further de-noise the data. Panels (c) and (d) in Supplementary Figure S2  show the FCC projections together with their averages (again black solid lines). The peaks are again 13 days apart.

Next, we seek an optimal alignment (or %perform a
\emph{registration})
%procedure that seeks an optimal alignment
of the curves which, separating horizontal and vertical variation, increases the statistical power of the analysis \cite{ramsay2005}.
Specifically, for each region, we estimate the horizontal shift (capped at a maximum of 20 days) that
%for each region as to minimize the 
minimizes the L2 distance 
%(i.e., $\int (u(t)-v(t))^2 dt$ for two generic curves $u$ and $v$)
$\int (R(t)-\tilde{R}(t))^2 dt$
between 
%each 
the $R_t$ curve and its FCC projection $\tilde{R}(t)$ 
%Notably, 
(FCC projections are an ideal alignment target here since they do not exhibit any horizontal variation; see Figure~S2).
%We fix the maximum possible shift at 20 days and 
We apply the estimated shifts to the $R_t$ as well as the $M_t$ curve in each region, preserving their time consistency. 
%within each region.
The shifted $R_t$ and $M_t$ curves restricted to the intersection of their new time domains are displayed in Figure~\ref{fig:fda_results} panels (a) and (b), respectively. The estimated shifts are shown in Supplementary Figure S2 panel (e). One can note how the registration procedure translates the epidemic curves aligning the $R_t$ peaks.

%We are now ready to perform a full
Finally, we apply function-on-function regression 
%model 
\cite{kokoszka2017introduction} to the registered curves to estimate the statistical effect of $M_t$ on $R_t$. 
%Mathematically, 
In symbols, we fit the model:
$$ y(t) = \alpha(t) + \int \beta_{mob}(s,t)M_t(s)ds + \epsilon(t)$$
where $y(t)$ is the shifted $R_t$ curve, $\alpha(t)$ is the intercept, $\epsilon(t)$ is the model error, and $M_t(s)$ is the shifted $M_t$.
Mobility flows are integrated over time and 
%its coefficient 
the effect coefficient $\beta_{mob}(s, t)$ is a surface that describes the association between $M_t$ at time $s$ and $R_t$ at time $t$. 
Given the previously estimated 13 days 
%effect 
%delay 
lag between the two sets of curves, the portion of surface we are most interested in is the curve identified by $\beta_{mob}(s,s+13)$, which represents the effect of mobility at time $s$ on
%$R_t$ 
transmission at time $s + 13$. 
%The functional regression model was fit using the R package \texttt{refund} \cite{goldsmith2016refund}.
Panels (c) and (d) in Figure~\ref{fig:fda_results} show the estimated surface $\hat \beta_{mob}(s,t)$ and the curve $\hat \beta_{mob}(s,s+13)$ with a $95\%$ confidence band \cite{goldsmith2013corrected}, respectively.
%This last curve clearly detects 
This suggests a strong positive effect of mobility on %transmission 
transmissibility in March and April, and a second period of significant positive association in June (in these 
%regions 
intervals of the domain the confidence bands are clearly above 0). The effect seems to be non-significant  %(i.e., the confidence bands contain 0) 
(confidence  bands comprising 0) at the end of the time domain -- which, because of 
%after 
the alignment process and the 13 days 
%delay 
lag is ante-dated to the beginning of September.

%At the end of the domain the effects of $M_t$ is null. 
The share of the variability in transmission explained by mobility on our data is high,
%-- which is a functional generalization of the coefficient of determination --
approximately  $R^2 = 0.73$.
%, meaning that the regression explains a large part of the variability in $R_t$. 

To further validate our results we performed 
% two additional analyses. First, we add a scalar
an additional analysis adding a scalar
predictor ($pc1$) to the regression model \cite{boschi2020shapes}. 
This is a composite control, obtained as the first principal component of a set of covariates that may affect the epidemic in addition to mobility -- see Supplementary Figure S3 for a complete list.
% specifically: adults per family doctor, average beds per hospital, average students per classroom, average employees per firm, and average members per household. 
Introducing this control variable does not change our inference on the effect of mobility flows (see Supplementary Figure S3). The total $R^2$ of the joint model is again $0.73$, while the the partial $R^2$ of mobility and pc1 are 0.24 and 0.01, respectively -- this indicates how the effect of mobility is considerably stronger than the one of the control variable.  
%
%{\red TB: SHOULD WE REMOVE PROVINCES COMPLETELY?
A a second form  of validation, we repeated our analysis using the $120$ Italian provinces.

This finer spatial resolution increases our sample size, and at the same time produces more idiosyncratic curves. 
$R_t$ is highly variable across provinces along the entire time domain considered. 
$M_t$ is highly variable across provinces too, but especially during the summer, reflecting differences between tourist destinations and other areas, with increased and depressed mobility flows, respectively. Notwithstanding these differences, function-on-function regression produces results consistent with those at the level of regions.
%The last row of {\red supplemental Figure S3} displays the regression results. Once again, the estimated effect of mobility is consistent with our main finding: also at a province level
$\hat\beta_{mob}(s,s+13)$ displays a strong positive effect in March and April and 
%an effect very close to 0 
a much weaker effect during the summer, with $R^2 = 0.53$.

\begin{figure}
% \begin{center}
% {\small \bf unshifted curves 2nd solution}
% \end{center}
\centering
% \includegraphics[width=0.4\linewidth]{./img/fda_img/rt_smoothed.pdf}
% \includegraphics[width=0.4\linewidth]{./img/fda_img/trips_smoothed.pdf} \\
% \includegraphics[width=0.4\linewidth]{./img/fda_img/rt_proj.pdf}
% \includegraphics[width=0.4\linewidth]{./img/fda_img/trips_proj.pdf} \\
%\vspace{-0.4cm}
\includegraphics[width=0.8\linewidth]{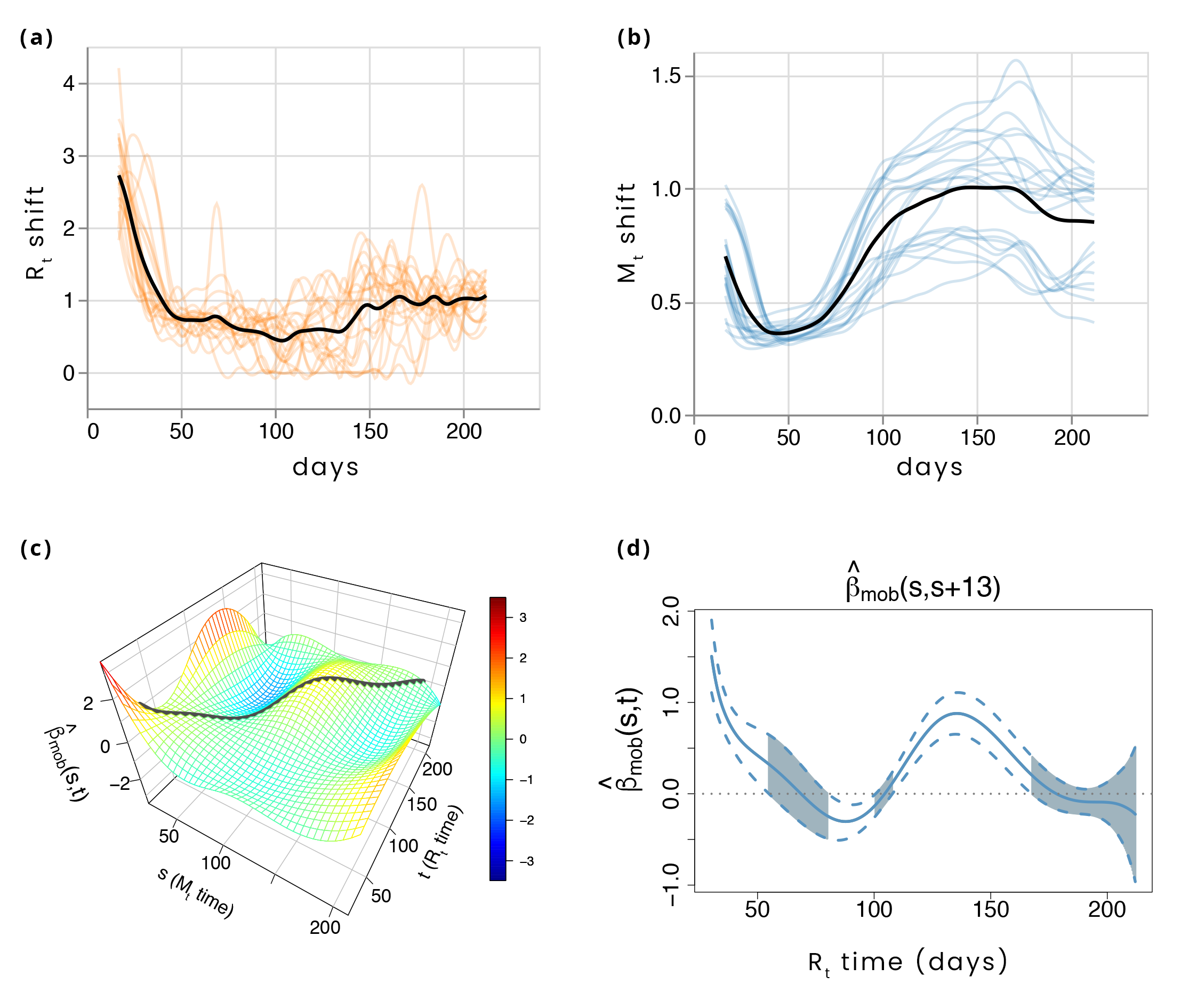}
\caption{\footnotesize Functional Data Analysis characterization of transmission and mobility.
% Panels in the first and second rows show smoothed $R_t$ and $M_t$ curves and their leading covariance component projections, respectively. 
% Thick black lines are the functional averages of the sets of curves. 
% 
Panels (a) and (b) display $R_t$ and $M_t$ smoothed registered curves restricted to the intersection of their new time domains. Thick black lines are the functional averages of the sets of curves. After the alignment the time domain is expressed in days, since the horizontal translation changes the calendar dates for each region. Day 1 corresponds to March 9th.
Panels in the bottom row show the estimated effect surface $\hat \beta(s,t)$ and the curve $\hat \beta(s,s+13)$ obtained along it (black ``cut'' along the surface) with a $95\%$ confidence band. The grey shaded areas represents the parts of the domain where the confidence bands contain 0. 
$\hat \beta(s,s+13)$ suggests a strong positive effect of mobility on transmission in March and April, and a second period of significant positive association in June. 
The share of the variability in transmission explained by mobility on our data is $R^2 = 0.73$.}
\label{fig:fda_results}
\end{figure}

\section{
%Conclusion
Conclusions}
Our combined analysis of mobility and epidemic data highlighted a striking relation between decreases in mobility flows and  net reproduction number, in all Italian regions, in 
%the 
a time interval of approximately one week (from March 11th till March 16th) %during the transition between the two mobility modalities
when  the country transitioned between pre- and post-lockdown mobility regimes. 
During this week, the two curves gracefully overlap -- 
%at the end of this week, the country has reached a new “stable” mobility regime, approximately at 40\% of the pre-lockdown level
until mobility stabilizes at about 40\% of its pre-lockdown level. 
The two curves exhibit the same pattern in most of the regions and provinces (see Supplementary Figure S1), with minimal temporal lags. 
We call this phenomenon, represented schematically in Figure 2e, the “epi-mob” pattern. Mobility (the blue curve) 
%is 
depicts a switch between the two 
%very different levels, before and during the lockdown, 
regimes, with an exponential fall from pre- to post-lockdown levels. 
%the first to the second; the
The net reproduction number 
%epidemics is a peaked distribution
(the orange curve) displays a  peak, with an exponential growth and fall,
%overlapping with the switch during the fall.
and overlaps with the switch during the latter.
 
%The presumable 
Effectiveness of the lockdown for the containment of the epidemics is further substantiated by the positive association we found between delays in mobility reduction and
total number of infected individuals per 100K inhabitants up to mid-May 2020. 
We can also quantify the time needed to “switch off” mobility in the country and the time needed to bring the net reproduction number below $1$ (approximately $1$ week).

Notice that $R_t$ continues to slowly decrease during lockdown, and also that at the beginning of Phase 2 (lockdown exit), when mobility begins to raise again, for most of the regions $R_t$ does not jump into a new uncontrolled growth for several months. 

While we do not perform any analysis that may elucidate the role of other factors, this may be due to other non-pharmaceutical interventions taking hold in the country, including increased compliance with the use personal protective equipment and social distancing (compared to the pre-lockdown phase). 

We have also shown, through functional data analysis, that the typical delay between the $M_t$ peak and the $R_t$ peak is around 2 weeks, both for regions and provinces.
This allowed us to create a regression model which shows a strong positive effect of mobility on transmission in March, April, and June, both at regional and provincial level.

In conclusion, we believe that this study demonstrates the value of digital mobility data, a detailed proxy of human behavior available every day in real time, for refining our understanding of the dynamics of the epidemics, evaluating the effectiveness of policy choices, and monitoring the unfolding of the epidemics in the coming months.

\begin{addendum}
 \item This work has been partially funded by the EU H2020 project SoBigData++ grant n.\ 871042. 
 We thank the former Italian ministry of Innovation Paola Pisano, together with Paolo De Rosa and the colleagues of the working group ``Big Data and AI for policy'' of the Italian data-driven task force for the COVID-19 epidemics\footnote{https://innovazione.gov.it/big-data-ai-for-policy}, for supporting our work. 
 \item[Competing interest] The authors declare no competing interests.
 \item[Code availability] The code to reproduce the analysis can be requested via email to luca.pappalardo@isti.cnr.it
 \item[Author contributions] PC made data managing, data preparation and wrote the paper; LP coordinated the work, made plot design, and wrote the paper; SR and DF made data managing and visualization; TB made the functional data analysis experiments and plots; FG coordinated the work and wrote the paper; FC coordinated the functional data analysis experiments and wrote the paper; DP coordinated the work and wrote the paper; GG and SM provided feedback on the results and helped interpret them; Vodafone provided the data and made data preparation; Authors affiliated to ISS and FBK computed the $R_t$ indicator.
\end{addendum}
\newpage
%\printbibliography
\bibliographystyle{naturemag}
\bibliography{biblio}

\begin{thebibliography}{10}
\expandafter\ifx\csname url\endcsname\relax
  \def\url#1{\texttt{#1}}\fi
\expandafter\ifx\csname urlprefix\endcsname\relax\def\urlprefix{URL }\fi
\providecommand{\bibinfo}[2]{#2}
\providecommand{\eprint}[2][]{\url{#2}}

\bibitem{perra2020nonpharmaceutical}
\bibinfo{author}{Perra, N.}
\newblock \bibinfo{title}{Non-pharmaceutical interventions during the covid-19
  pandemic: a rapid review} (\bibinfo{year}{2020}).
\newblock \eprint{2012.15230}.

\bibitem{bo2020effectiveness}
\bibinfo{author}{Bo, Y.} \emph{et~al.}
\newblock \bibinfo{title}{Effectiveness of non-pharmaceutical interventions on
  covid-19 transmission in 190 countries from 23 january to 13 april 2020}.
\newblock \emph{\bibinfo{journal}{International Journal of Infectious
  Diseases}} \textbf{\bibinfo{volume}{102}}, \bibinfo{pages}{247 -- 253}
  (\bibinfo{year}{2021}).
\newblock
  \urlprefix\url{http://www.sciencedirect.com/science/article/pii/S1201971220322700}.

\bibitem{haug2020ranking}
\bibinfo{author}{Haug, N.} \emph{et~al.}
\newblock \bibinfo{title}{Ranking the effectiveness of worldwide covid-19
  government interventions}.
\newblock \emph{\bibinfo{journal}{Nature human behaviour}}
  \bibinfo{pages}{1--10} (\bibinfo{year}{2020}).

\bibitem{blondel2015survey}
\bibinfo{author}{Blondel, V.~D.}, \bibinfo{author}{Decuyper, A.} \&
  \bibinfo{author}{Krings, G.}
\newblock \bibinfo{title}{A survey of results on mobile phone datasets
  analysis}.
\newblock \emph{\bibinfo{journal}{EPJ Data Science}}
  \textbf{\bibinfo{volume}{4}}, \bibinfo{pages}{10} (\bibinfo{year}{2015}).
\newblock \urlprefix\url{https://doi.org/10.1140/epjds/s13688-015-0046-0}.

\bibitem{barbosa2018human}
\bibinfo{author}{Barbosa, H.} \emph{et~al.}
\newblock \bibinfo{title}{Human mobility: Models and applications}.
\newblock \emph{\bibinfo{journal}{Physics Reports}}
  \textbf{\bibinfo{volume}{734}}, \bibinfo{pages}{1--74}
  (\bibinfo{year}{2018}).

\bibitem{luca2020deep}
\bibinfo{author}{Luca, M.}, \bibinfo{author}{Barlacchi, G.},
  \bibinfo{author}{Lepri, B.} \& \bibinfo{author}{Pappalardo, L.}
\newblock \bibinfo{title}{Deep learning for human mobility: a survey on data
  and models} (\bibinfo{year}{2020}).
\newblock \eprint{2012.02825}.

\bibitem{giannotti2008mobility}
\bibinfo{author}{Giannotti, F.} \& \bibinfo{author}{Pedreschi, D.}
\newblock \emph{\bibinfo{title}{Mobility, data mining and privacy: Geographic
  knowledge discovery}} (\bibinfo{publisher}{Springer Science \& Business
  Media}, \bibinfo{year}{2008}).

\bibitem{wang2018applying}
\bibinfo{author}{Wang, Z.}, \bibinfo{author}{He, S.~Y.} \&
  \bibinfo{author}{Leung, Y.}
\newblock \bibinfo{title}{Applying mobile phone data to travel behaviour
  research: A literature review}.
\newblock \emph{\bibinfo{journal}{Travel Behaviour and Society}}
  \textbf{\bibinfo{volume}{11}}, \bibinfo{pages}{141 -- 155}
  (\bibinfo{year}{2018}).
\newblock
  \urlprefix\url{http://www.sciencedirect.com/science/article/pii/S2214367X17300224}.

\bibitem{gonzalez2008understanding}
\bibinfo{author}{Gonzalez, M.~C.}, \bibinfo{author}{Hidalgo, C.~A.} \&
  \bibinfo{author}{Barabasi, A.-L.}
\newblock \bibinfo{title}{Understanding individual human mobility patterns}.
\newblock \emph{\bibinfo{journal}{nature}} \textbf{\bibinfo{volume}{453}},
  \bibinfo{pages}{779--782} (\bibinfo{year}{2008}).

\bibitem{pappalardo2015returners}
\bibinfo{author}{Pappalardo, L.} \emph{et~al.}
\newblock \bibinfo{title}{Returners and explorers dichotomy in human mobility}.
\newblock \emph{\bibinfo{journal}{Nature Communications}}
  \textbf{\bibinfo{volume}{6}}, \bibinfo{pages}{8166} (\bibinfo{year}{2015}).
\newblock \urlprefix\url{https://doi.org/10.1038/ncomms9166}.

\bibitem{alessandretti2018evidence}
\bibinfo{author}{Alessandretti, L.}, \bibinfo{author}{Sapiezynski, P.},
  \bibinfo{author}{Sekara, V.}, \bibinfo{author}{Lehmann, S.} \&
  \bibinfo{author}{Baronchelli, A.}
\newblock \bibinfo{title}{Evidence for a conserved quantity in human mobility}.
\newblock \emph{\bibinfo{journal}{Nature Human Behaviour}}
  \textbf{\bibinfo{volume}{2}}, \bibinfo{pages}{485--491}
  (\bibinfo{year}{2018}).

\bibitem{simini2012universal}
\bibinfo{author}{Simini, F.}, \bibinfo{author}{Gonz{\'a}lez, M.~C.},
  \bibinfo{author}{Maritan, A.} \& \bibinfo{author}{Barab{\'a}si, A.-L.}
\newblock \bibinfo{title}{A universal model for mobility and migration
  patterns}.
\newblock \emph{\bibinfo{journal}{Nature}} \textbf{\bibinfo{volume}{484}},
  \bibinfo{pages}{96--100} (\bibinfo{year}{2012}).

\bibitem{pappalardo2016analytical}
\bibinfo{author}{Pappalardo, L.} \emph{et~al.}
\newblock \bibinfo{title}{An analytical framework to nowcast well-being using
  mobile phone data}.
\newblock \emph{\bibinfo{journal}{International Journal of Data Science and
  Analytics}} \textbf{\bibinfo{volume}{2}}, \bibinfo{pages}{75--92}
  (\bibinfo{year}{2016}).

\bibitem{eagle2009eigenbehaviors}
\bibinfo{author}{Eagle, N.} \& \bibinfo{author}{Pentland, A.~S.}
\newblock \bibinfo{title}{Eigenbehaviors: Identifying structure in routine}.
\newblock \emph{\bibinfo{journal}{Behavioral Ecology and Sociobiology}}
  \textbf{\bibinfo{volume}{63}}, \bibinfo{pages}{1057--1066}
  (\bibinfo{year}{2009}).

\bibitem{song2010limits}
\bibinfo{author}{Song, C.}, \bibinfo{author}{Qu, Z.}, \bibinfo{author}{Blumm,
  N.} \& \bibinfo{author}{Barab{\'a}si, A.-L.}
\newblock \bibinfo{title}{Limits of predictability in human mobility}.
\newblock \emph{\bibinfo{journal}{Science}} \textbf{\bibinfo{volume}{327}},
  \bibinfo{pages}{1018--1021} (\bibinfo{year}{2010}).

\bibitem{barbosa2020uncovering}
\bibinfo{author}{Barbosa, H.} \emph{et~al.}
\newblock \bibinfo{title}{Uncovering the socioeconomic facets of human
  mobility}.
\newblock \emph{\bibinfo{journal}{arXiv preprint arXiv:2012.00838}}
  (\bibinfo{year}{2020}).

\bibitem{gabrielli2015city}
\bibinfo{author}{Gabrielli, L.}, \bibinfo{author}{Furletti, B.},
  \bibinfo{author}{Trasarti, R.}, \bibinfo{author}{Giannotti, F.} \&
  \bibinfo{author}{Pedreschi, D.}
\newblock \bibinfo{title}{City users' classification with mobile phone data}.
\newblock In \emph{\bibinfo{booktitle}{2015 IEEE International Conference on
  Big Data (Big Data)}}, \bibinfo{pages}{1007--1012} (\bibinfo{year}{2015}).

\bibitem{deville2014dynamic}
\bibinfo{author}{Deville, P.} \emph{et~al.}
\newblock \bibinfo{title}{Dynamic population mapping using mobile phone data}.
\newblock \emph{\bibinfo{journal}{Proceedings of the National Academy of
  Sciences}} \textbf{\bibinfo{volume}{111}}, \bibinfo{pages}{15888--15893}
  (\bibinfo{year}{2014}).
\newblock \urlprefix\url{https://www.pnas.org/content/111/45/15888}.
\newblock \eprint{https://www.pnas.org/content/111/45/15888.full.pdf}.

\bibitem{douglass2015high}
\bibinfo{author}{Douglass, R.~W.}, \bibinfo{author}{Meyer, D.~A.},
  \bibinfo{author}{Ram, M.}, \bibinfo{author}{Rideout, D.} \&
  \bibinfo{author}{Song, D.}
\newblock \bibinfo{title}{High resolution population estimates from
  telecommunications data}.
\newblock \emph{\bibinfo{journal}{EPJ Data Science}}
  \textbf{\bibinfo{volume}{4}}, \bibinfo{pages}{4} (\bibinfo{year}{2015}).

\bibitem{hankaew2019inferring}
\bibinfo{author}{Hankaew, S.} \emph{et~al.}
\newblock \bibinfo{title}{Inferring and modeling migration flows using mobile
  phone network data}.
\newblock \emph{\bibinfo{journal}{IEEE Access}} \textbf{\bibinfo{volume}{7}},
  \bibinfo{pages}{164746--164758} (\bibinfo{year}{2019}).

\bibitem{balzotti2018understanding}
\bibinfo{author}{Balzotti, C.}, \bibinfo{author}{Bragagnini, A.},
  \bibinfo{author}{Briani, M.} \& \bibinfo{author}{Cristiani, E.}
\newblock \bibinfo{title}{Understanding human mobility flows from aggregated
  mobile phone data}.
\newblock \emph{\bibinfo{journal}{IFAC-PapersOnLine}}
  \textbf{\bibinfo{volume}{51}}, \bibinfo{pages}{25 -- 30}
  (\bibinfo{year}{2018}).
\newblock
  \urlprefix\url{http://www.sciencedirect.com/science/article/pii/S2405896318307213}.
\newblock \bibinfo{note}{15th IFAC Symposium on Control in Transportation
  Systems CTS 2018}.

\bibitem{bonnel2018origin}
\bibinfo{author}{Bonnel, P.}, \bibinfo{author}{Fekih, M.} \&
  \bibinfo{author}{Smoreda, Z.}
\newblock \bibinfo{title}{Origin-destination estimation using mobile network
  probe data}.
\newblock \emph{\bibinfo{journal}{Transportation Research Procedia}}
  \textbf{\bibinfo{volume}{32}}, \bibinfo{pages}{69--81}
  (\bibinfo{year}{2018}).

\bibitem{liang2020using}
\bibinfo{author}{Liang, L.}, \bibinfo{author}{Shrestha, R.},
  \bibinfo{author}{Ghosh, S.} \& \bibinfo{author}{Webb, P.}
\newblock \bibinfo{title}{Using mobile phone data helps estimate
  community-level food insecurity: Findings from a multi-year panel study in
  nepal}.
\newblock \emph{\bibinfo{journal}{PLOS ONE}} \textbf{\bibinfo{volume}{15}},
  \bibinfo{pages}{1--16} (\bibinfo{year}{2020}).
\newblock \urlprefix\url{https://doi.org/10.1371/journal.pone.0241791}.

\bibitem{voukelatou2020measuring}
\bibinfo{author}{Voukelatou, V.} \emph{et~al.}
\newblock \bibinfo{title}{Measuring objective and subjective well-being:
  dimensions and data sources}.
\newblock \emph{\bibinfo{journal}{International Journal of Data Science and
  Analytics}} \bibinfo{pages}{1--31} (\bibinfo{year}{2020}).

\bibitem{eagle2010network}
\bibinfo{author}{Eagle, N.}, \bibinfo{author}{Macy, M.} \&
  \bibinfo{author}{Claxton, R.}
\newblock \bibinfo{title}{Network diversity and economic development}.
\newblock \emph{\bibinfo{journal}{Science}} \textbf{\bibinfo{volume}{328}},
  \bibinfo{pages}{1029--1031} (\bibinfo{year}{2010}).

\bibitem{frias2012relationship}
\bibinfo{author}{Frias-Martinez, V.} \& \bibinfo{author}{Virseda, J.}
\newblock \bibinfo{title}{On the relationship between socio-economic factors
  and cell phone usage}.
\newblock In \emph{\bibinfo{booktitle}{Proceedings of the fifth international
  conference on information and communication technologies and development}},
  \bibinfo{pages}{76--84} (\bibinfo{year}{2012}).

\bibitem{blumenstock2015predicting}
\bibinfo{author}{Blumenstock, J.}, \bibinfo{author}{Cadamuro, G.} \&
  \bibinfo{author}{On, R.}
\newblock \bibinfo{title}{Predicting poverty and wealth from mobile phone
  metadata}.
\newblock \emph{\bibinfo{journal}{Science}} \textbf{\bibinfo{volume}{350}},
  \bibinfo{pages}{1073--1076} (\bibinfo{year}{2015}).

\bibitem{lai2019exploring}
\bibinfo{author}{Lai, S.} \emph{et~al.}
\newblock \bibinfo{title}{Exploring the use of mobile phone data for national
  migration statistics}.
\newblock \emph{\bibinfo{journal}{Palgrave communications}}
  \textbf{\bibinfo{volume}{5}}, \bibinfo{pages}{1--10} (\bibinfo{year}{2019}).

\bibitem{chi2020general}
\bibinfo{author}{Chi, G.}, \bibinfo{author}{Lin, F.}, \bibinfo{author}{Chi, G.}
  \& \bibinfo{author}{Blumenstock, J.}
\newblock \bibinfo{title}{A general approach to detecting migration events in
  digital trace data}.
\newblock \emph{\bibinfo{journal}{PloS one}} \textbf{\bibinfo{volume}{15}},
  \bibinfo{pages}{e0239408} (\bibinfo{year}{2020}).

\bibitem{blumenstock2012inferring}
\bibinfo{author}{Blumenstock, J.~E.}
\newblock \bibinfo{title}{Inferring patterns of internal migration from mobile
  phone call records: evidence from rwanda}.
\newblock \emph{\bibinfo{journal}{Information Technology for Development}}
  \textbf{\bibinfo{volume}{18}}, \bibinfo{pages}{107--125}
  (\bibinfo{year}{2012}).

\bibitem{hughes2016inferring}
\bibinfo{author}{Hughes, C.} \emph{et~al.}
\newblock \bibinfo{title}{Inferring migrations: Traditional methods and new
  approaches based on mobile phone, social media, and other big data:
  Feasibility study on inferring (labour) mobility and migration in the
  european union from big data and social media data}  (\bibinfo{year}{2016}).

\bibitem{demontoye2018privacy}
\bibinfo{author}{de~Montjoye, Y.-A.} \emph{et~al.}
\newblock \bibinfo{title}{On the privacy-conscientious use of mobile phone
  data}.
\newblock \emph{\bibinfo{journal}{Scientific Data}}
  \textbf{\bibinfo{volume}{5}}, \bibinfo{pages}{180286} (\bibinfo{year}{2018}).
\newblock \urlprefix\url{https://doi.org/10.1038/sdata.2018.286}.

\bibitem{pellungrini2017data}
\bibinfo{author}{Pellungrini, R.}, \bibinfo{author}{Pappalardo, L.},
  \bibinfo{author}{Pratesi, F.} \& \bibinfo{author}{Monreale, A.}
\newblock \bibinfo{title}{A data mining approach to assess privacy risk in
  human mobility data}.
\newblock \emph{\bibinfo{journal}{ACM Trans. Intell. Syst. Technol.}}
  \textbf{\bibinfo{volume}{9}} (\bibinfo{year}{2017}).
\newblock \urlprefix\url{https://doi.org/10.1145/3106774}.

\bibitem{fiore2019privacy}
\bibinfo{author}{Fiore, M.} \emph{et~al.}
\newblock \bibinfo{title}{Privacy in trajectory micro-data publishing : a
  survey}.
\newblock \emph{\bibinfo{journal}{arXiv: Cryptography and Security}}
  (\bibinfo{year}{2019}).

\bibitem{demontjoye2013unique}
\bibinfo{author}{de~Montjoye, Y.-A.}, \bibinfo{author}{Hidalgo, C.~A.},
  \bibinfo{author}{Verleysen, M.} \& \bibinfo{author}{Blondel, V.~D.}
\newblock \bibinfo{title}{Unique in the crowd: The privacy bounds of human
  mobility}.
\newblock \emph{\bibinfo{journal}{Scientific Reports}}
  \textbf{\bibinfo{volume}{3}} (\bibinfo{year}{2013}).

\bibitem{oliver2020mobile}
\bibinfo{author}{Oliver, N.} \emph{et~al.}
\newblock \bibinfo{title}{Mobile phone data for informing public health actions
  across the covid-19 pandemic life cycle}.
\newblock \emph{\bibinfo{journal}{Science Advances}}
  \textbf{\bibinfo{volume}{6}} (\bibinfo{year}{2020}).
\newblock
  \urlprefix\url{https://advances.sciencemag.org/content/6/23/eabc0764}.
\newblock
  \eprint{https://advances.sciencemag.org/content/6/23/eabc0764.full.pdf}.

\bibitem{buckee2020aggregated}
\bibinfo{author}{Buckee, C.~O.} \emph{et~al.}
\newblock \bibinfo{title}{Aggregated mobility data could help fight covid-19.}
\newblock \emph{\bibinfo{journal}{Science (New York, NY)}}
  \textbf{\bibinfo{volume}{368}}, \bibinfo{pages}{145} (\bibinfo{year}{2020}).

\bibitem{finger2016mobile}
\bibinfo{author}{Finger, F.} \emph{et~al.}
\newblock \bibinfo{title}{Mobile phone data highlights the role of mass
  gatherings in the spreading of cholera outbreaks}.
\newblock \emph{\bibinfo{journal}{Proceedings of the National Academy of
  Sciences}} \textbf{\bibinfo{volume}{113}}, \bibinfo{pages}{6421--6426}
  (\bibinfo{year}{2016}).

\bibitem{tizzoni2014use}
\bibinfo{author}{Tizzoni, M.} \emph{et~al.}
\newblock \bibinfo{title}{On the use of human mobility proxies for modeling
  epidemics}.
\newblock \emph{\bibinfo{journal}{PLoS Comput Biol}}
  \textbf{\bibinfo{volume}{10}}, \bibinfo{pages}{e1003716}
  (\bibinfo{year}{2014}).

\bibitem{wesolowski2012quantifying}
\bibinfo{author}{Wesolowski, A.} \emph{et~al.}
\newblock \bibinfo{title}{Quantifying the impact of human mobility on malaria}.
\newblock \emph{\bibinfo{journal}{Science}} \textbf{\bibinfo{volume}{338}},
  \bibinfo{pages}{267--270} (\bibinfo{year}{2012}).

\bibitem{bengtsson2015using}
\bibinfo{author}{Bengtsson, L.} \emph{et~al.}
\newblock \bibinfo{title}{Using mobile phone data to predict the spatial spread
  of cholera}.
\newblock \emph{\bibinfo{journal}{Scientific reports}}
  \textbf{\bibinfo{volume}{5}}, \bibinfo{pages}{8923} (\bibinfo{year}{2015}).

\bibitem{kraemer2020effect}
\bibinfo{author}{Kraemer, M.~U.} \emph{et~al.}
\newblock \bibinfo{title}{The effect of human mobility and control measures on
  the covid-19 epidemic in china}.
\newblock \emph{\bibinfo{journal}{Science}} \textbf{\bibinfo{volume}{368}},
  \bibinfo{pages}{493--497} (\bibinfo{year}{2020}).

\bibitem{bonato2020mobile}
\bibinfo{author}{Bonato, P.} \emph{et~al.}
\newblock \bibinfo{title}{Mobile phone data analytics against the covid-19
  epidemics in italy: flow diversity and local job markets during the national
  lockdown} (\bibinfo{year}{2020}).
\newblock \eprint{2004.11278}.

\bibitem{pullano2020population}
\bibinfo{author}{Pullano, G.}, \bibinfo{author}{Valdano, E.},
  \bibinfo{author}{Scarpa, N.}, \bibinfo{author}{Rubrichi, S.} \&
  \bibinfo{author}{Colizza, V.}
\newblock \bibinfo{title}{Population mobility reductions during covid-19
  epidemic in france under lockdown}.
\newblock \emph{\bibinfo{journal}{medRxiv}}  (\bibinfo{year}{2020}).

\bibitem{lai2020effect}
\bibinfo{author}{Lai, S.} \emph{et~al.}
\newblock \bibinfo{title}{Effect of non-pharmaceutical interventions to contain
  covid-19 in china}  (\bibinfo{year}{2020}).

\bibitem{liautaud2020fever}
\bibinfo{author}{Liautaud, P.}, \bibinfo{author}{Huybers, P.} \&
  \bibinfo{author}{Santillana, M.}
\newblock \bibinfo{title}{Fever and mobility data indicate social distancing
  has reduced incidence of communicable disease in the united states}
  (\bibinfo{year}{2020}).
\newblock \eprint{2004.09911}.

\bibitem{badr2020association}
\bibinfo{author}{Badr, H.~S.} \emph{et~al.}
\newblock \bibinfo{title}{Association between mobility patterns and covid-19
  transmission in the usa: a mathematical modelling study}.
\newblock \emph{\bibinfo{journal}{The Lancet Infectious Diseases}}
  (\bibinfo{year}{2020}).

\bibitem{coven2020disparities}
\bibinfo{author}{Coven, J.} \& \bibinfo{author}{Gupta, A.}
\newblock \bibinfo{title}{Disparities in mobility responses to covid-19}.
\newblock \bibinfo{type}{Tech. Rep.}, \bibinfo{institution}{NYU Stern Working
  Paper} (\bibinfo{year}{2020}).

\bibitem{gozzi2020estimating}
\bibinfo{author}{Gozzi, N.} \emph{et~al.}
\newblock \bibinfo{title}{Estimating the effect of social inequalities in the
  mitigation of covid-19 across communities in santiago de chile}.
\newblock \emph{\bibinfo{journal}{medRxiv}}  (\bibinfo{year}{2020}).

\bibitem{bakker2020effect}
\bibinfo{author}{Bakker, M.}, \bibinfo{author}{Berke, A.},
  \bibinfo{author}{Groh, M.}, \bibinfo{author}{Pentland, A.} \&
  \bibinfo{author}{Moro, E.}
\newblock \bibinfo{title}{Effect of social distancing measures in the new york
  city metropolitan area}.
\newblock \bibinfo{type}{Tech. Rep.}, \bibinfo{institution}{Massachusetts
  Institute of Technology} (\bibinfo{year}{2020}).

\bibitem{jia2020population}
\bibinfo{author}{Jia, J.~S.} \emph{et~al.}
\newblock \bibinfo{title}{Population flow drives spatio-temporal distribution
  of covid-19 in china}.
\newblock \emph{\bibinfo{journal}{Nature}} \bibinfo{pages}{1--5}
  (\bibinfo{year}{2020}).

\bibitem{kissler2020reductions}
\bibinfo{author}{Kissler, S.} \emph{et~al.}
\newblock \bibinfo{title}{Reductions in commuting mobility predict geographic
  differences in sars-cov-2 prevalence in new york city}
  (\bibinfo{year}{2020}).

\bibitem{gibbs2020changing}
\bibinfo{author}{Gibbs, H.} \emph{et~al.}
\newblock \bibinfo{title}{Changing travel patterns in china during the early
  stages of the covid-19 pandemic}.
\newblock \emph{\bibinfo{journal}{medRxiv}}  (\bibinfo{year}{2020}).

\bibitem{kang2020multiscale}
\bibinfo{author}{Kang, Y.} \emph{et~al.}
\newblock \bibinfo{title}{Multiscale dynamic human mobility flow dataset in the
  us during the covid-19 epidemic}.
\newblock \emph{\bibinfo{journal}{Scientific data}}
  \textbf{\bibinfo{volume}{7}}, \bibinfo{pages}{1--13} (\bibinfo{year}{2020}).

\bibitem{pepe2020covid}
\bibinfo{author}{Pepe, E.} \emph{et~al.}
\newblock \bibinfo{title}{Covid-19 outbreak response, a dataset to assess
  mobility changes in italy following national lockdown}.
\newblock \emph{\bibinfo{journal}{Scientific data}}
  \textbf{\bibinfo{volume}{7}}, \bibinfo{pages}{1--7} (\bibinfo{year}{2020}).

\bibitem{iqbal2014development}
\bibinfo{author}{Iqbal, M.~S.}, \bibinfo{author}{Choudhury, C.~F.},
  \bibinfo{author}{Wang, P.} \& \bibinfo{author}{Gonz{\'a}lez, M.~C.}
\newblock \bibinfo{title}{Development of origin--destination matrices using
  mobile phone call data}.
\newblock \emph{\bibinfo{journal}{Transportation Research Part C: Emerging
  Technologies}} \textbf{\bibinfo{volume}{40}}, \bibinfo{pages}{63--74}
  (\bibinfo{year}{2014}).

\bibitem{bachir2019inferring}
\bibinfo{author}{Bachir, D.}, \bibinfo{author}{Khodabandelou, G.},
  \bibinfo{author}{Gauthier, V.}, \bibinfo{author}{El~Yacoubi, M.} \&
  \bibinfo{author}{Puchinger, J.}
\newblock \bibinfo{title}{Inferring dynamic origin-destination flows by
  transport mode using mobile phone data}.
\newblock \emph{\bibinfo{journal}{Transportation Research Part C: Emerging
  Technologies}} \textbf{\bibinfo{volume}{101}}, \bibinfo{pages}{254--275}
  (\bibinfo{year}{2019}).

\bibitem{riccardo2020epidemiological}
\bibinfo{author}{Riccardo, F.} \emph{et~al.}
\newblock \bibinfo{title}{Epidemiological characteristics of covid-19 cases in
  italy and estimates of the reproductive numbers one month into the epidemic}.
\newblock \emph{\bibinfo{journal}{medRxiv}}  (\bibinfo{year}{2020}).

\bibitem{who2014ebola}
\bibinfo{author}{Team, W. E.~R.}
\newblock \bibinfo{title}{Ebola virus disease in west africa—the first 9
  months of the epidemic and forward projections}.
\newblock \emph{\bibinfo{journal}{New England Journal of Medicine}}
  \textbf{\bibinfo{volume}{371}}, \bibinfo{pages}{1481--1495}
  (\bibinfo{year}{2014}).

\bibitem{cori2013new}
\bibinfo{author}{Cori, A.}, \bibinfo{author}{Ferguson, N.~M.},
  \bibinfo{author}{Fraser, C.} \& \bibinfo{author}{Cauchemez, S.}
\newblock \bibinfo{title}{A new framework and software to estimate time-varying
  reproduction numbers during epidemics}.
\newblock \emph{\bibinfo{journal}{American journal of epidemiology}}
  \textbf{\bibinfo{volume}{178}}, \bibinfo{pages}{1505--1512}
  (\bibinfo{year}{2013}).

\bibitem{liu2018measurability}
\bibinfo{author}{Liu, Q.-H.} \emph{et~al.}
\newblock \bibinfo{title}{Measurability of the epidemic reproduction number in
  data-driven contact networks}.
\newblock \emph{\bibinfo{journal}{Proceedings of the National Academy of
  Sciences}} \textbf{\bibinfo{volume}{115}}, \bibinfo{pages}{12680--12685}
  (\bibinfo{year}{2018}).

\bibitem{nishiura2020serial}
\bibinfo{author}{Nishiura, H.}, \bibinfo{author}{Linton, N.~M.} \&
  \bibinfo{author}{Akhmetzhanov, A.~R.}
\newblock \bibinfo{title}{Serial interval of novel coronavirus (covid-19)
  infections}.
\newblock \emph{\bibinfo{journal}{International journal of infectious
  diseases}}  (\bibinfo{year}{2020}).

\bibitem{wu2020estimating}
\bibinfo{author}{Wu, J.~T.} \emph{et~al.}
\newblock \bibinfo{title}{Estimating clinical severity of covid-19 from the
  transmission dynamics in wuhan, china}.
\newblock \emph{\bibinfo{journal}{Nature Medicine}}
  \textbf{\bibinfo{volume}{26}}, \bibinfo{pages}{506--510}
  (\bibinfo{year}{2020}).

\bibitem{li2020early}
\bibinfo{author}{Li, Q.} \emph{et~al.}
\newblock \bibinfo{title}{Early transmission dynamics in wuhan, china, of novel
  coronavirus--infected pneumonia}.
\newblock \emph{\bibinfo{journal}{New England Journal of Medicine}}
  (\bibinfo{year}{2020}).

\bibitem{ramsay2005}
\bibinfo{author}{Ramsay, J.~O.} \& \bibinfo{author}{Silverman, B.~W.}
\newblock \emph{\bibinfo{title}{Functional data analysis}}
  (\bibinfo{publisher}{Springer}, \bibinfo{year}{2005}), \bibinfo{edition}{2}
  edn.

\bibitem{craven1978smoothing}
\bibinfo{author}{Craven, P.} \& \bibinfo{author}{Wahba, G.}
\newblock \bibinfo{title}{Smoothing noisy data with spline functions}.
\newblock \emph{\bibinfo{journal}{Numerische mathematik}}
  \textbf{\bibinfo{volume}{31}}, \bibinfo{pages}{377--403}
  (\bibinfo{year}{1978}).

\bibitem{boschi2018covariance}
\bibinfo{author}{Boschi, T.}, \bibinfo{author}{Chiaromonte, F.},
  \bibinfo{author}{Secchi, P.} \& \bibinfo{author}{Li, B.}
\newblock \bibinfo{title}{Covariance based low-dimensional registration for
  function-on-function regression}  (\bibinfo{year}{2018}).

\bibitem{kokoszka2017introduction}
\bibinfo{author}{Kokoszka, P.} \& \bibinfo{author}{Reimherr, M.}
\newblock \emph{\bibinfo{title}{Introduction to functional data analysis}}
  (\bibinfo{publisher}{CRC Press}, \bibinfo{year}{2017}).

\bibitem{goldsmith2013corrected}
\bibinfo{author}{Goldsmith, J.}, \bibinfo{author}{Greven, S.} \&
  \bibinfo{author}{Crainiceanu, C.}
\newblock \bibinfo{title}{Corrected confidence bands for functional data using
  principal components}.
\newblock \emph{\bibinfo{journal}{Biometrics}} \textbf{\bibinfo{volume}{69}},
  \bibinfo{pages}{41--51} (\bibinfo{year}{2013}).

\bibitem{boschi2020shapes}
\bibinfo{author}{Boschi, T.}, \bibinfo{author}{Di~Iorio, J.},
  \bibinfo{author}{Testa, L.}, \bibinfo{author}{Cremona, M.~A.} \&
  \bibinfo{author}{Chiaromonte, F.}
\newblock \bibinfo{title}{The shapes of an epidemic: using functional data
  analysis to characterize covid-19 in italy}.
\newblock \emph{\bibinfo{journal}{arXiv preprint arXiv:2008.04700}}
  (\bibinfo{year}{2020}).

\end{thebibliography}

\newpage
\section*{Supplementary Figures}

\begin{figure}
\centering
\includegraphics[width=0.8\linewidth]{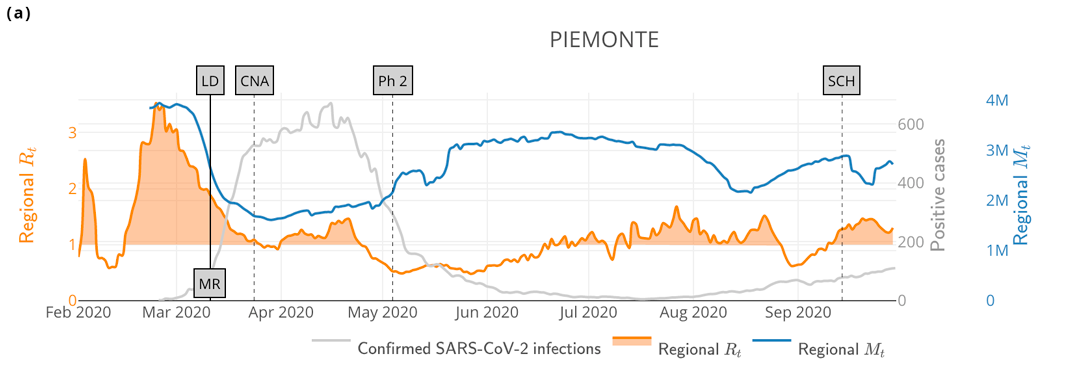}
\includegraphics[width=0.8\linewidth]{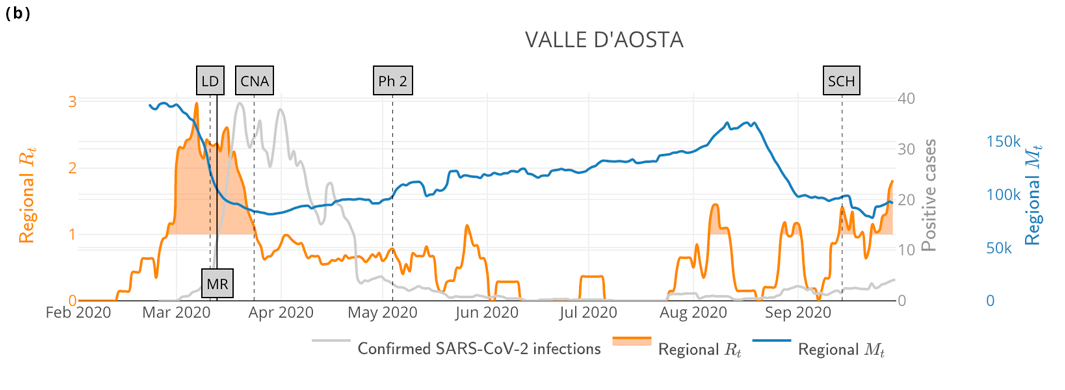}
\includegraphics[width=0.8\linewidth]{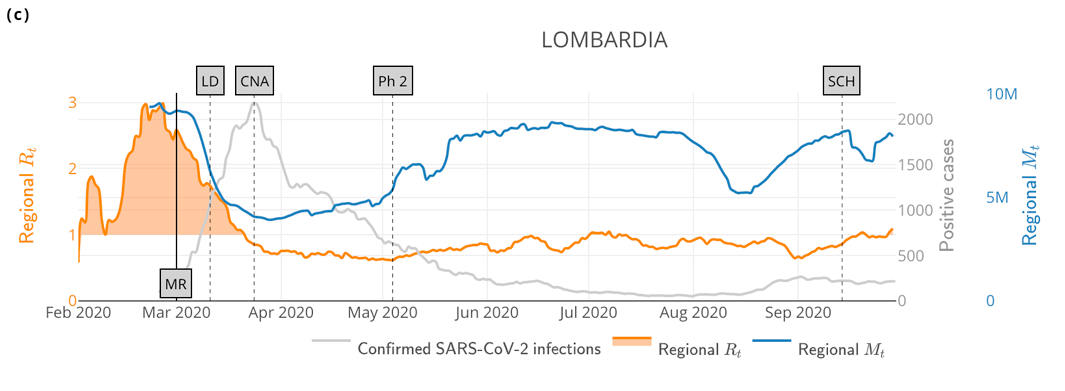}
\includegraphics[width=0.8\linewidth]{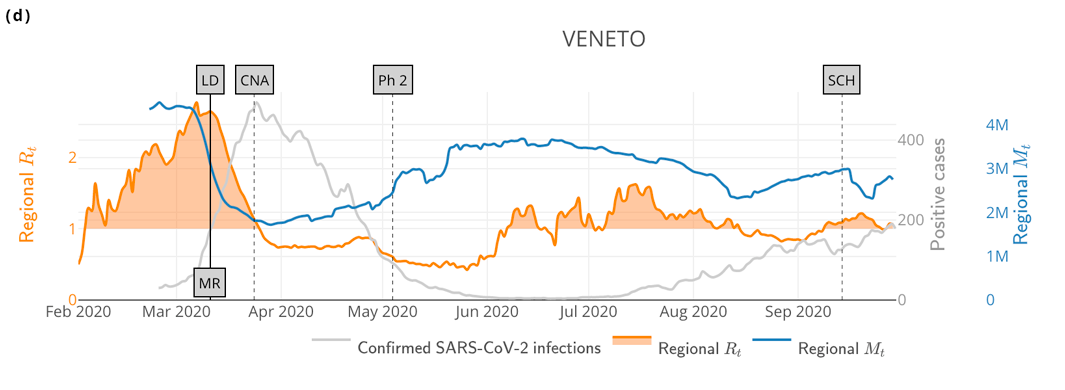}
\end{figure}

\begin{figure}
\centering
\vspace{-1.5cm}
\includegraphics[width=0.8\linewidth]{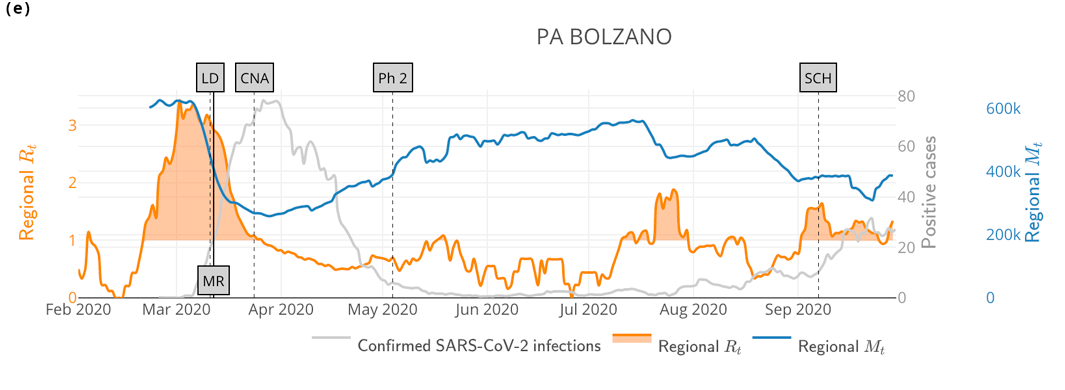}
\includegraphics[width=0.8\linewidth]{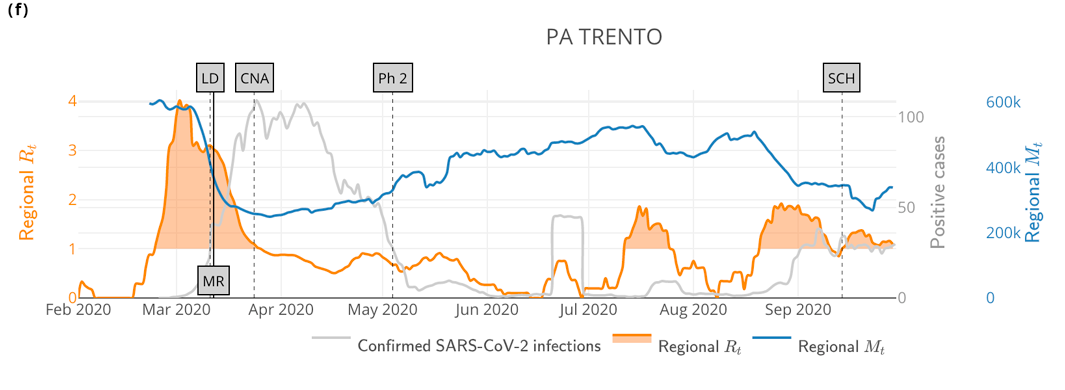}
\includegraphics[width=0.8\linewidth]{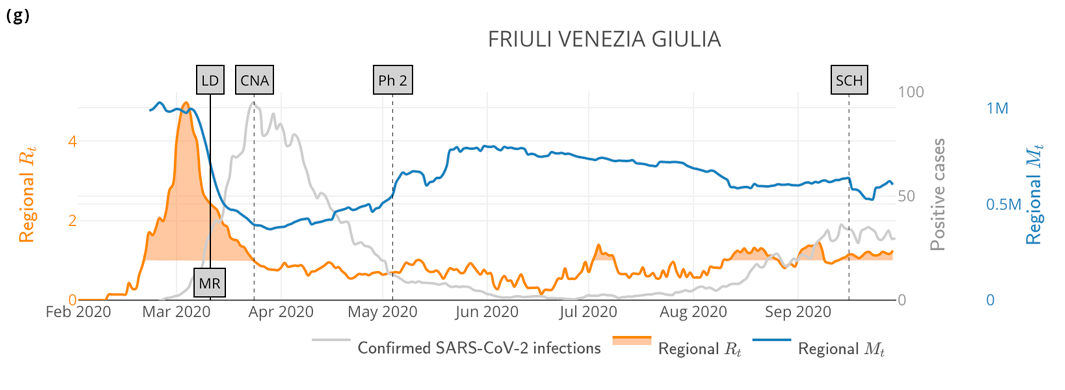}
\includegraphics[width=0.8\linewidth]{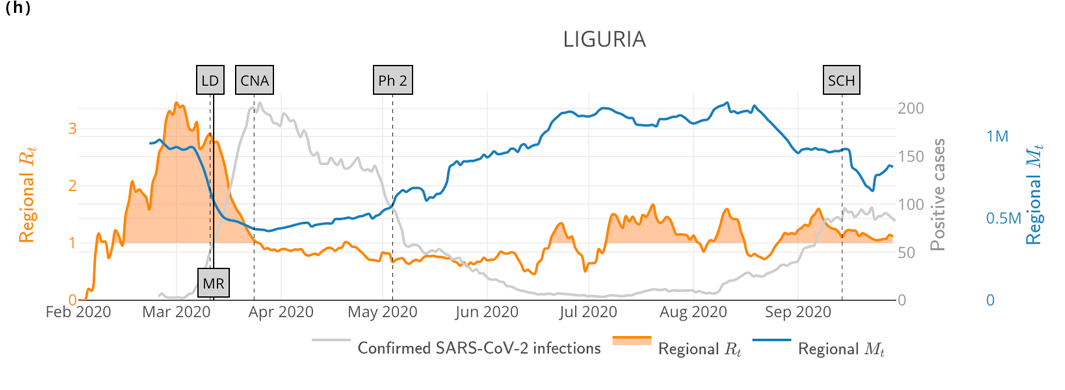}
\end{figure}

\begin{figure}
\centering
\vspace{-1.5cm}
\includegraphics[width=0.8\linewidth]{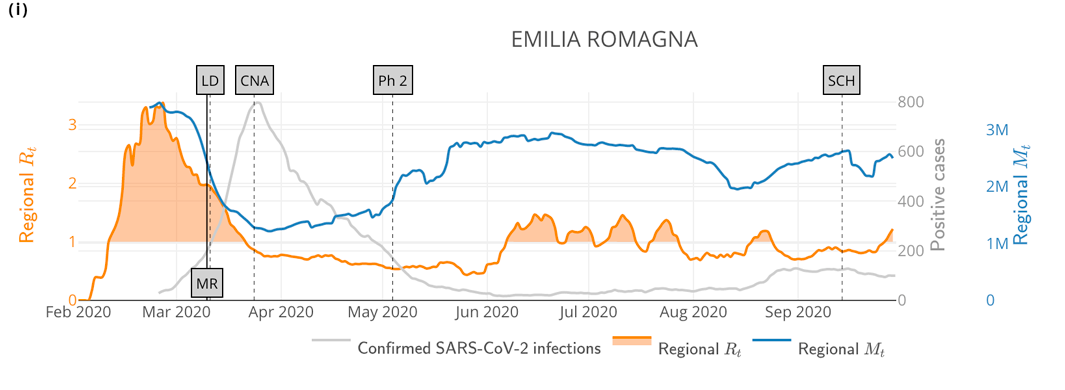}
\includegraphics[width=0.8\linewidth]{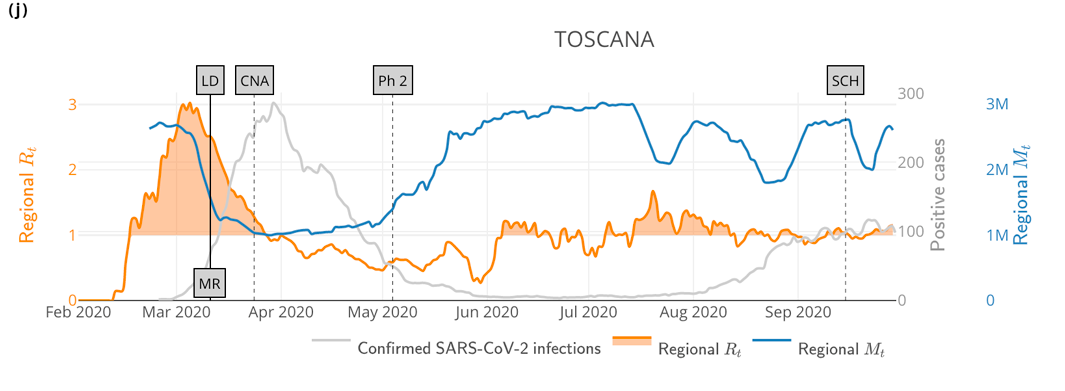}
\includegraphics[width=0.8\linewidth]{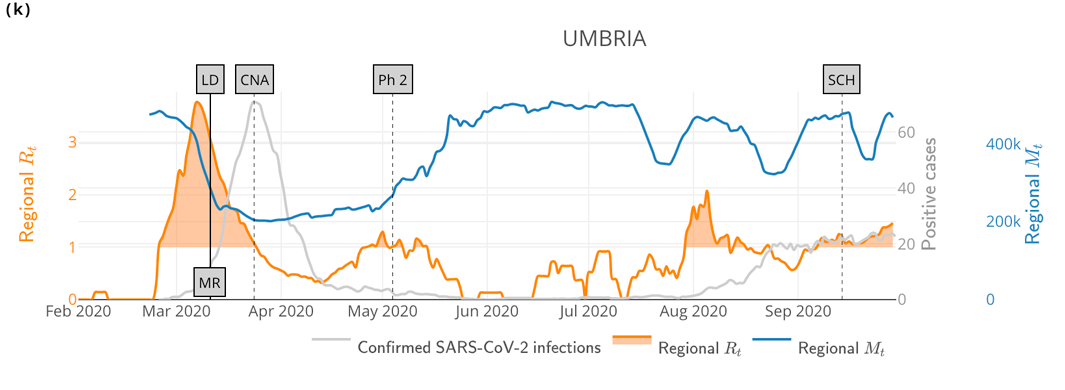}
\includegraphics[width=0.8\linewidth]{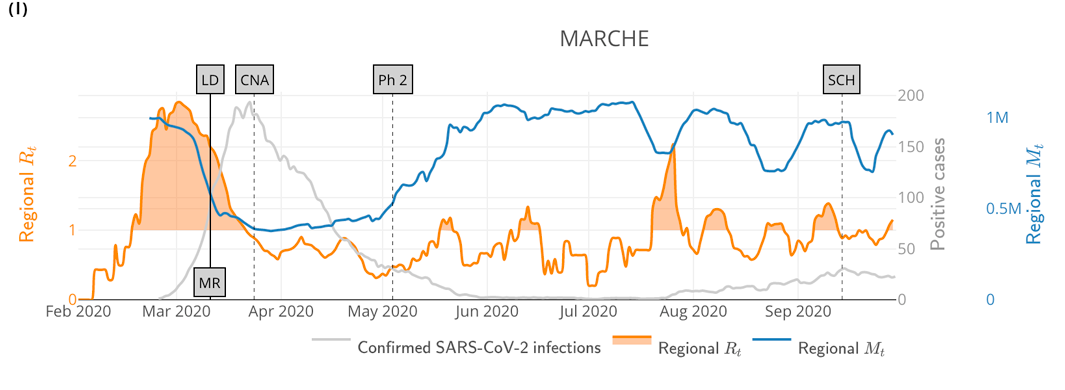}
\end{figure}

\begin{figure}
\centering
\includegraphics[width=0.8\linewidth]{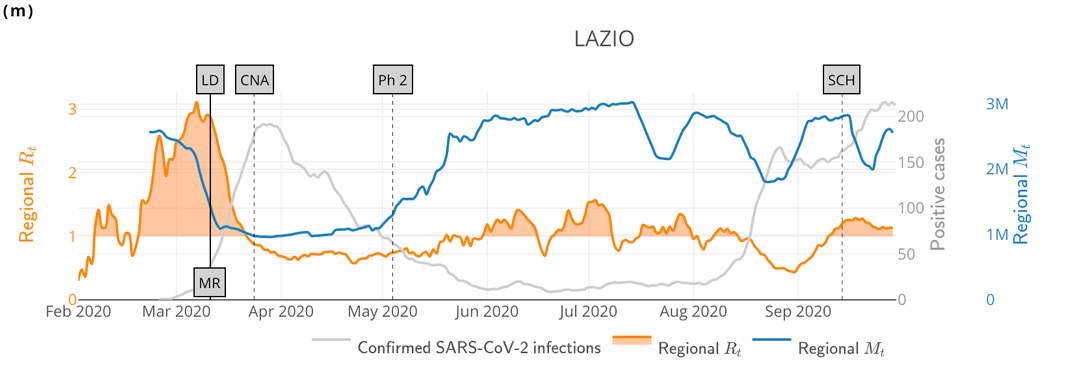}
\includegraphics[width=0.8\linewidth]{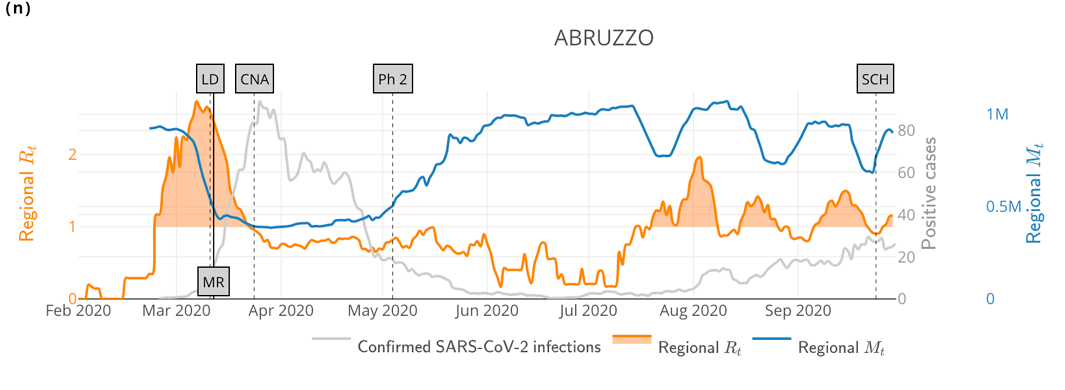}
\includegraphics[width=0.8\linewidth]{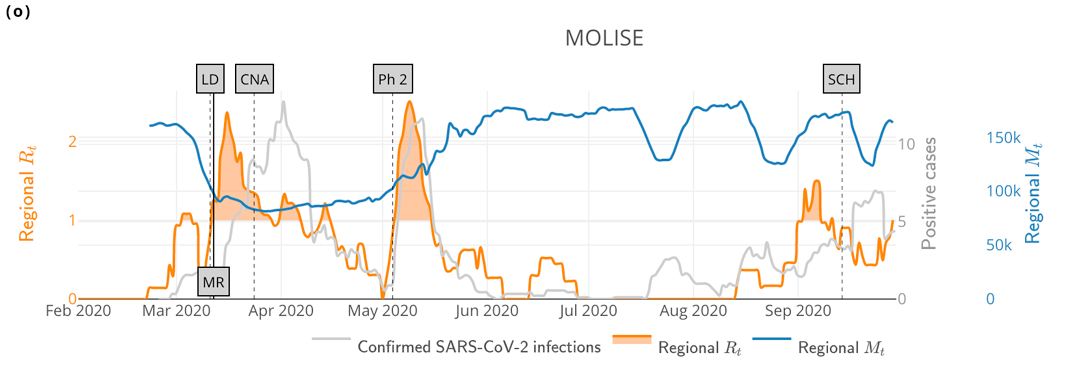}
\includegraphics[width=0.8\linewidth]{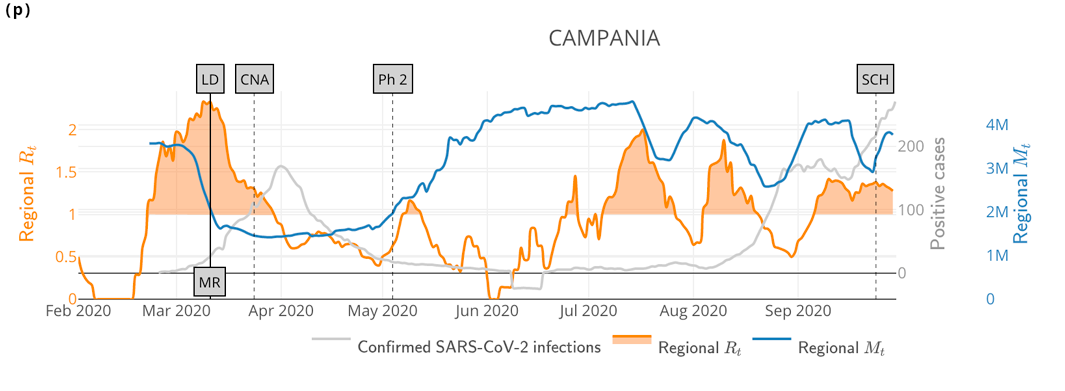}
\end{figure}

\begin{figure}
\centering
\includegraphics[width=0.8\linewidth]{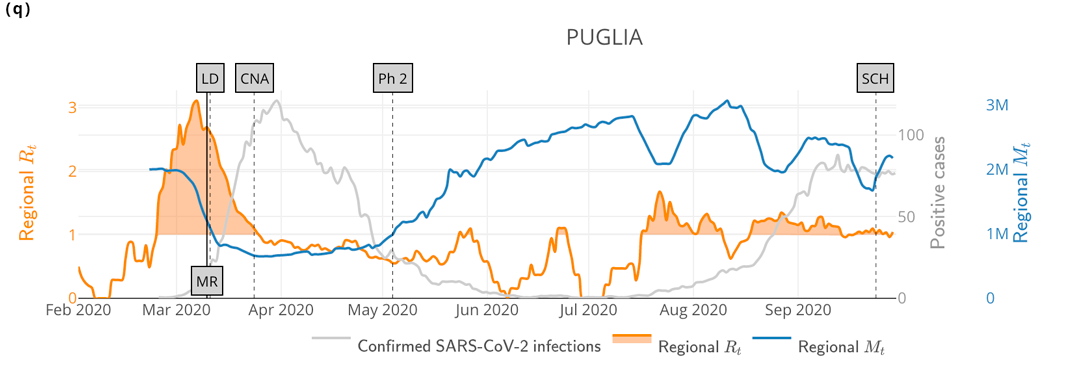}
\includegraphics[width=0.8\linewidth]{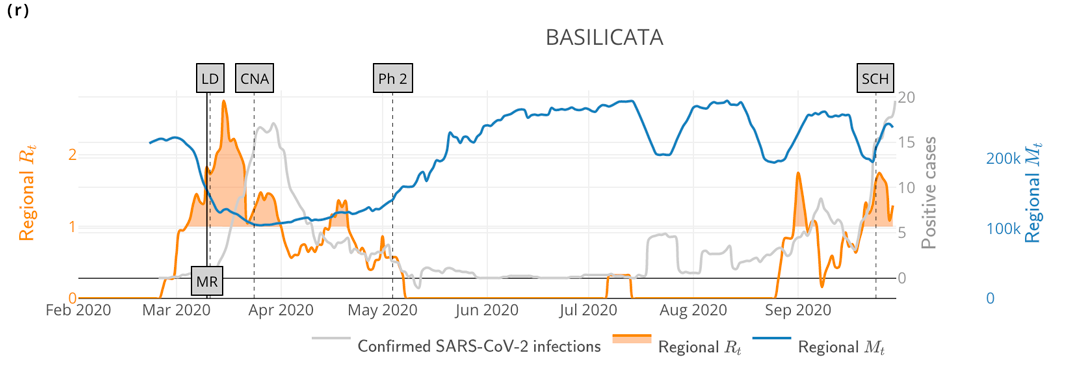}
\includegraphics[width=0.8\linewidth]{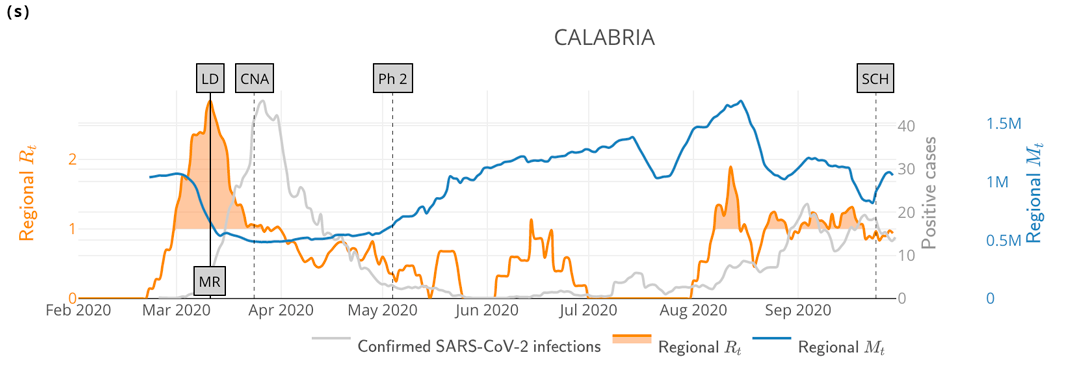}
\includegraphics[width=0.8\linewidth]{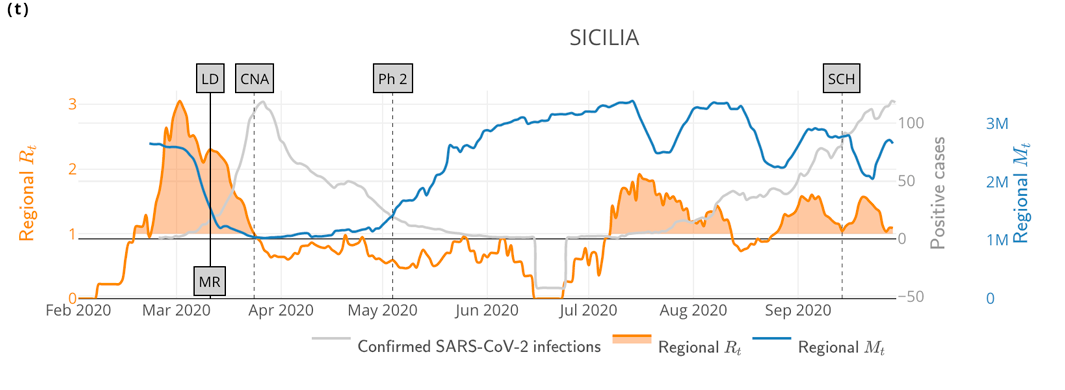}

\end{figure}

\begin{figure}
\centering
\includegraphics[width=0.8\linewidth]{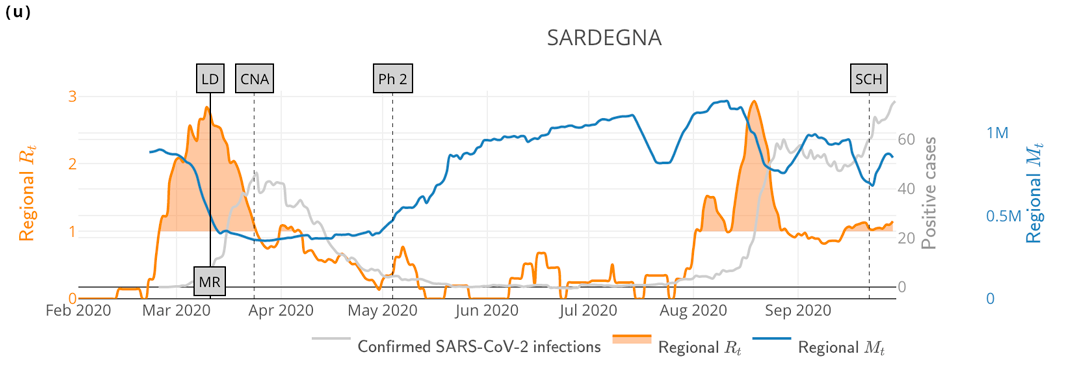}
\caption{Line chart representations of the $M_t$ (blue), $R_t$ (orange) and the number of positive cases (grey) series for all the Italian regions. In each chart the series are normalized  and projected on a dedicated axis for each curve, for an effective comparison of their evolution.}
\end{figure}

\begin{figure}
\centering
\includegraphics[width=1\linewidth]{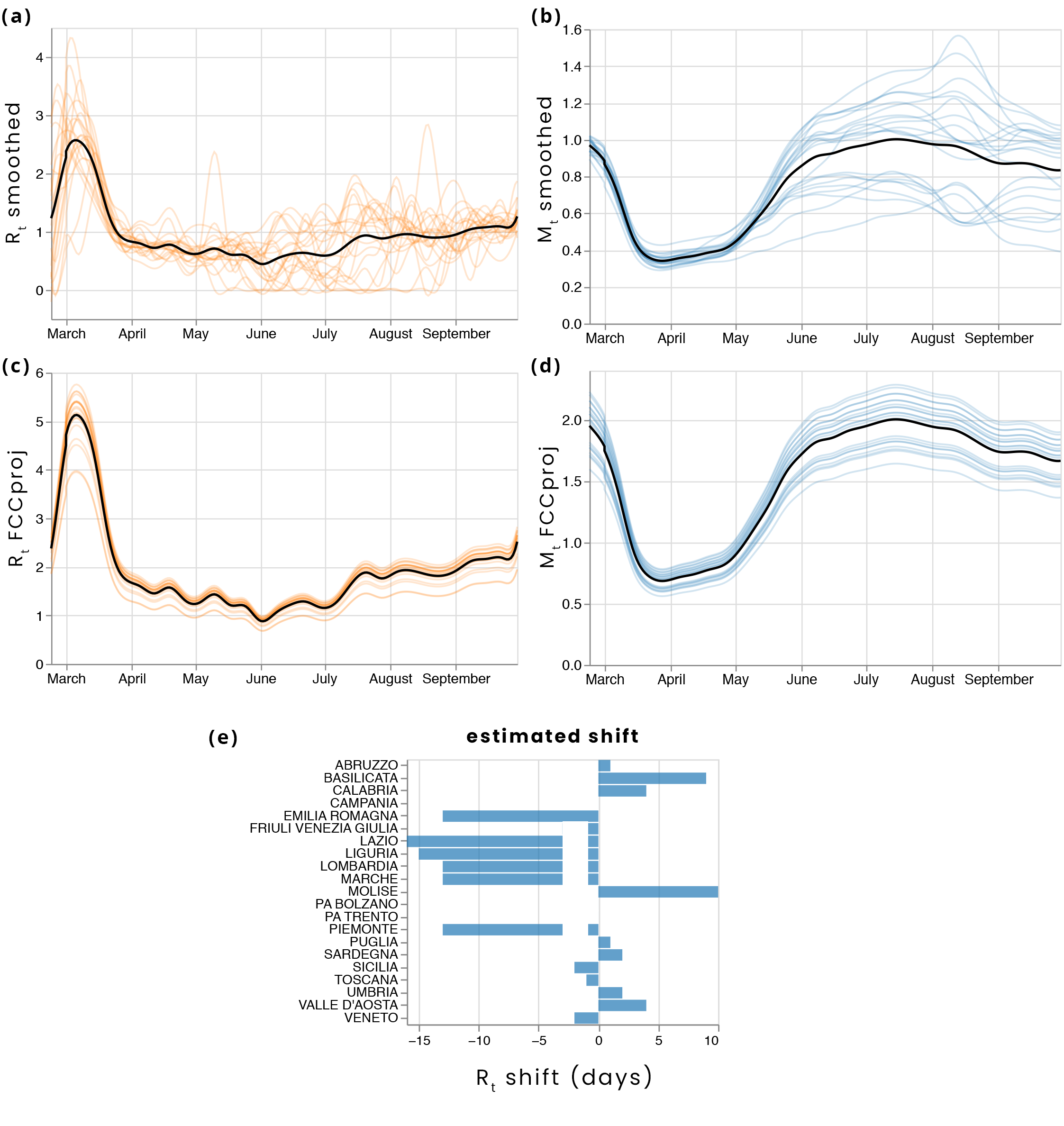}
\caption{Panels (a) and (b) show smoothed $R_t$ and $M_t$ curves. Panels (c) and (d) display their leading covariance component projections. 
Thick black lines are the functional averages of the sets of curves. 
Panel (e) depicts the estimated horizontal shifts by the registration procedure. 
}
\end{figure}

\begin{figure}
% \begin{center}
% {\small \bf unshifted curves 2nd solution}
% \end{center}
\centering
\includegraphics[width=0.32\linewidth]{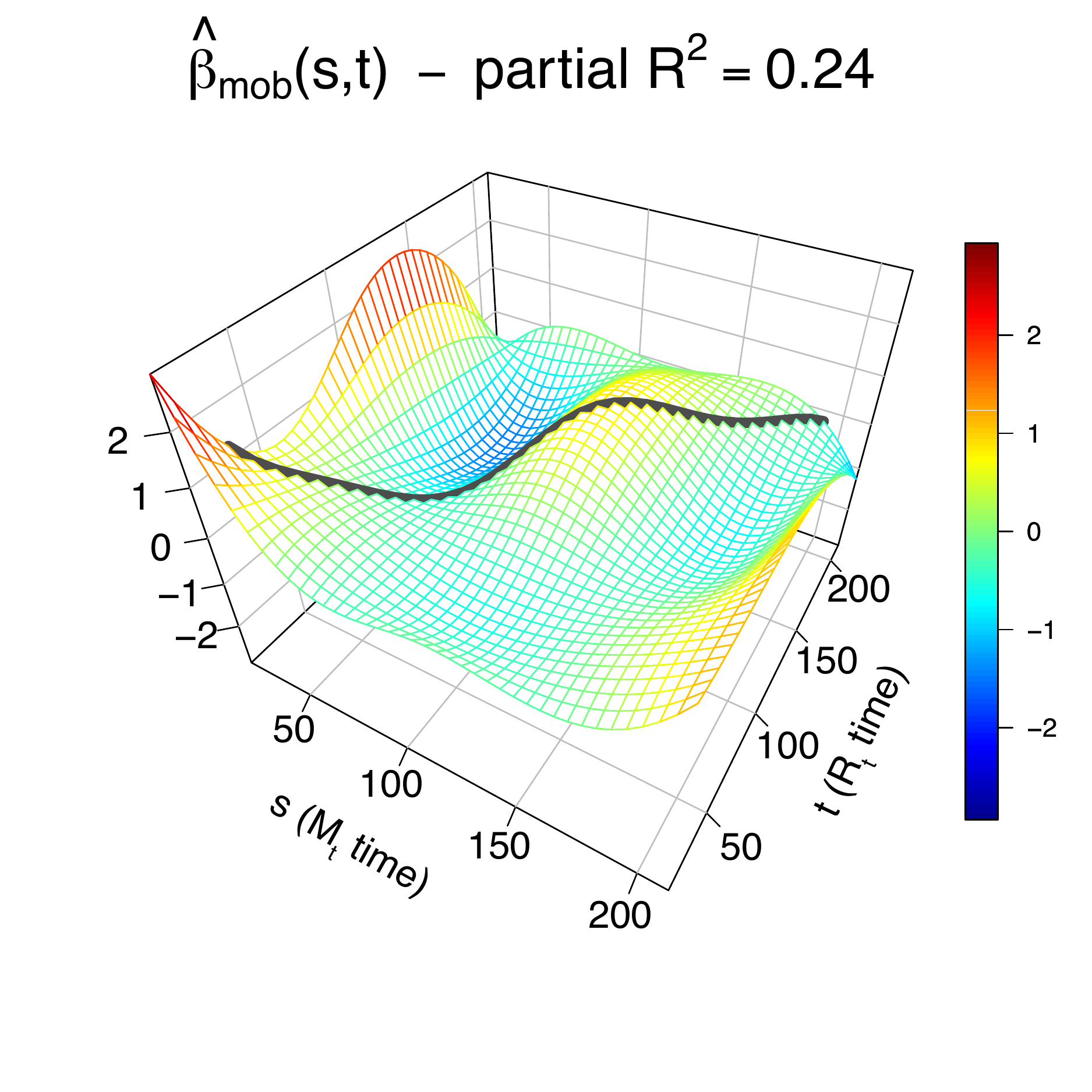}
\includegraphics[width=0.32\linewidth]{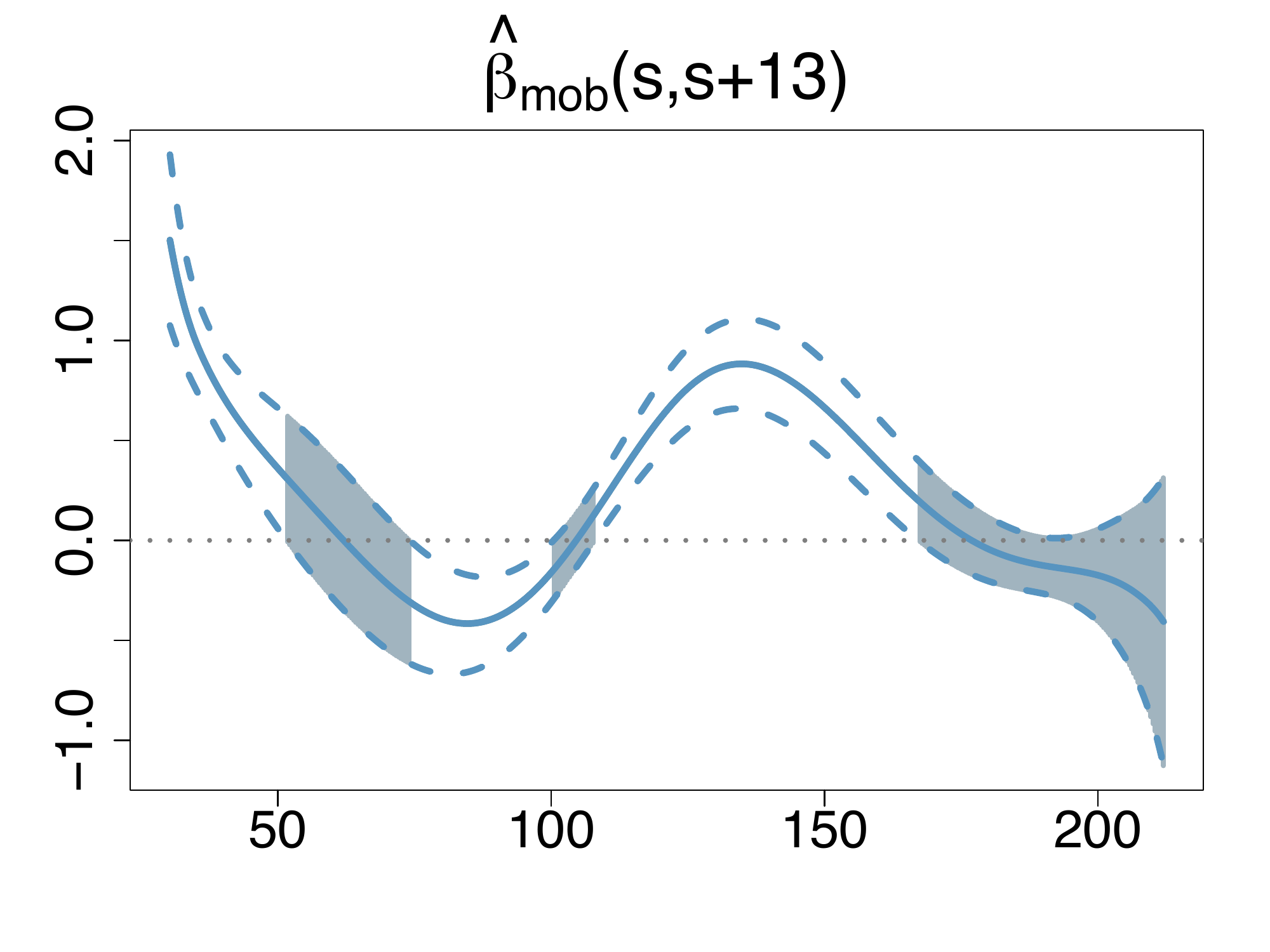}
\includegraphics[width=0.32\linewidth]{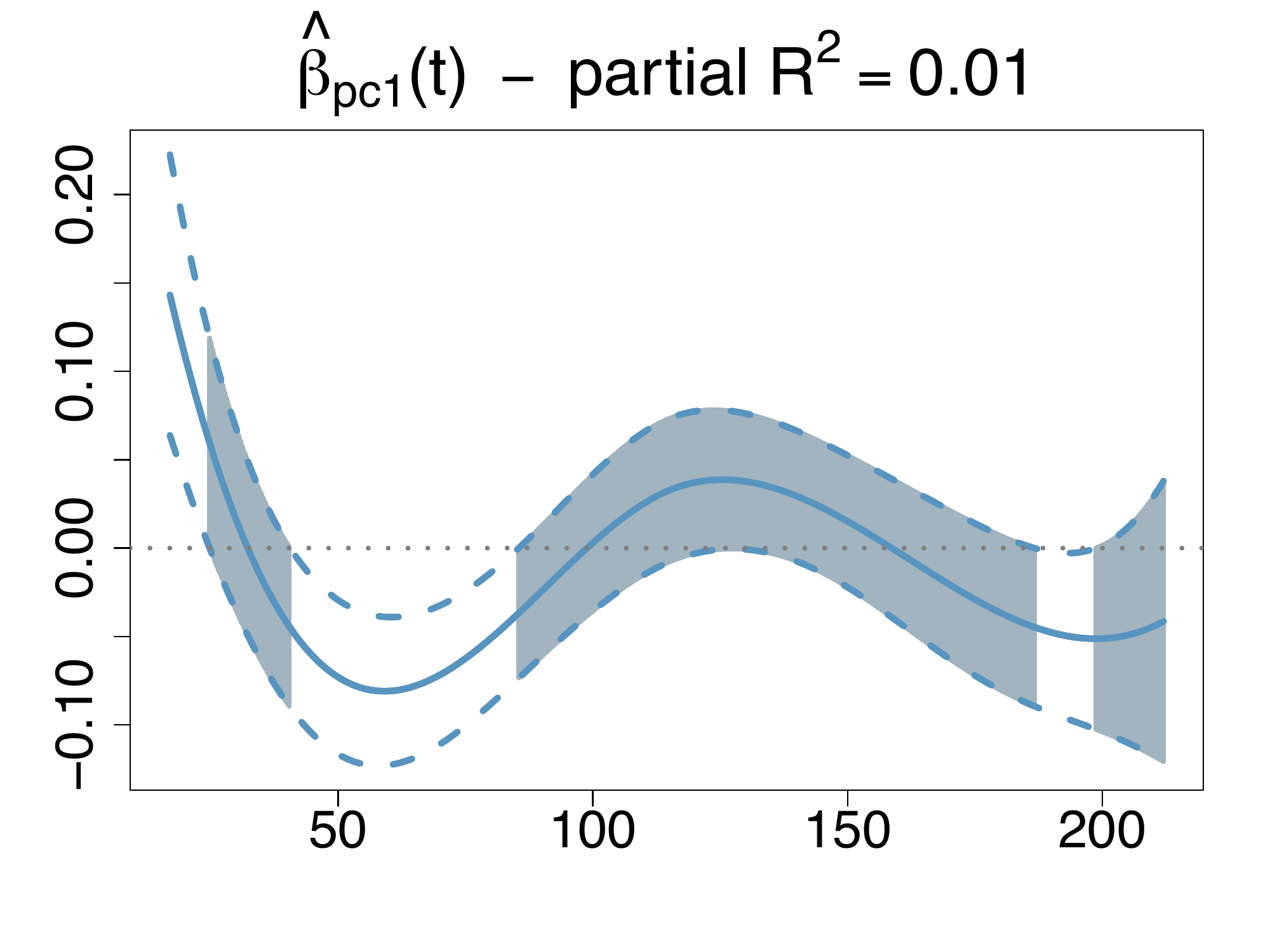}
\caption{Results of the joint functional regression model. Transmission curves are regressed against mobility curves and the scalar variable $pc1$.
This is a composite control, obtained as the first principal component of a set of covariates that may affect the epidemic in addition to mobility -- specifically: adults per family doctor, average beds per hospital, average students per classroom, average employees per firm, and average members per household.
The left and the center panels shows the estimated effect surface $\hat \beta_{mob}(s,t)$ for mobility and the curve $\hat \beta_{mob}(s,s+13)$ obtained along it (black ``cut'' along the surface) with a $95\%$ confidence band.
The right panel shows the estimated $pc1$ effect curve with a $95\%$ confidence band. 
The gray shaded areas correspond to the parts of the domain where the confidence bands include the 0. 
The time along the axis is expressed in days. The starting date after the alignment process is postponed to March 9th.
$\hat \beta_{mob}(s,s+13)$ suggests a strong positive effect of mobility on transmission in March and April, and a second period of significant positive association in June. $\hat \beta_{pc}(t)$ has an effect very close to 0 along with the all domain.
The total $R^2$ of the model is 0.73, while the partial $R^2$ of mobility and pc1 are 0.26 and 0.01, respectively.
% \textcolor{red}{[FC: TOBIA, RECAPITULATE HERE KEY RESULTS. THEY ARE ALREADY IN THE MAIN TEXT, BUT IT DOES NOT HURT TO ITERATE HERE]}}
}
\label{fig:suppl_regre}
\end{figure}

\end{document}